\documentstyle[aps,psfig]{revtex}

\newcommand{\be}{\begin{equation}}
\newcommand{\ee}{\end{equation}}
\newcommand{\bea}{\begin{eqnarray}}
\newcommand{\eea}{\end{eqnarray}}

\newcommand{\lsim}{\stackrel{\scriptscriptstyle
<}{\scriptscriptstyle \sim}}
\newcommand{\gsim}{\stackrel{\scriptscriptstyle
>}{\scriptscriptstyle \sim}}
\newcommand{\toh}{{\textstyle {1\over 2}}}
 \newcommand{\half}{{\textstyle {1\over 2}}}

\newcommand{\Eq}[1]{Eq.~(\ref{#1})}
\newcommand{\Eqs}[1]{Eqs.~(\ref{#1})}
\newcommand{\sect}[1]{section~\ref{#1}}
\newcommand{\fig}[1]{Fig.~\ref{#1}}
\newcommand{\sfrac}[2]{\mbox{$\textstyle {#1 \over #2}$}}
\newcommand{\sss}{\scriptscriptstyle }
\newcommand{\kb}{k_{\scriptscriptstyle B}}
\newcommand{\Tk}{T_{\scriptscriptstyle K}}
\newcommand{\Vk}{V_{\scriptscriptstyle K}}
\newcommand{\ve}{\varepsilon}

\newcommand{\nfl}{non-Fermi-liquid}
\newcommand{\nflr}{non-Fermi-liquid regime}

\begin{document}

\include{labelsII}

\draft

\title{The 2-Channel Kondo Model I: Review of Experimental
Evidence for its Realization in Metal Nanoconstrictions}

\author{Jan von Delft${}^{1,}$\cite{newadd},
D. C. Ralph${}^1$, R. A. Buhrman${}^2$,
S. K. Upadhyay${}^1$, R. N. Louie${}^1$,\\
A. W. W. Ludwig${}^3$, 
Vinay Ambegaokar${}^1$}
\address{
${}^1$Laboratory of Atomic and Solid State Physics, Cornell University,
Ithaca, NY 14853, USA \\
${}^2$School of Applied and Engineering Physics, Cornell University,
Ithaca, NY 14853, USA \\
${}^3$ Department of Physics,
University of California, Santa Barbara, CA 93106 , USA}

\twocolumn[\hsize\textwidth\columnwidth\hsize\csname%
@twocolumnfalse\endcsname
\date{February 4, 1997}
\maketitle 


\begin{abstract}
  Certain zero-bias anomalies (ZBAs) in the voltage, temperature and
  magnetic field dependence of the conductance $G(V,T,H)$ of quenched
  Cu point contacts have previously been interpreted to be due to
  non-magnetic 2-channel Kondo (2CK) scattering from near-degenerate
  atomic two-level tunneling systems (Ralph and Buhrman, 1992; Ralph
  {\em et al.}\/ 1994), and hence to represent an experimental
  realization of the non-Fermi-liquid physics of the $T=0$ fixed point
  of the 2-channel Kondo model.  In this, the first in a series of
  three papers (I,II,III) devoted to 2-channel Kondo physics, we
  present a comprehensive review of the quenched Cu ZBA experiments
  and their 2CK interpretation, including new results on ZBAs in
  constrictions made from Ti or from metallic glasses.  We first
  review the evidence that the ZBAs are due to electron scattering
  from stuctural defects that are not static, but possess internal
  dynamics.  In order to distinguish between several mechanisms
  proposed to explain the experiments, we then analyze the scaling
  properties of the conductance at low temperature and voltage and
  extract from the data a universal scaling function $\Gamma(v)$.  The
  theoretical calculation of the corresponding scaling function within
  the 2CK model is the subject of papers II and III.  The main
  conclusion of our work is that the properties of the ZBAs, and most
  notably their scaling behavior, are in good agreement with the 2CK
  model and clearly different from several other proposed mechanisms.
\end{abstract}
\vspace*{0.5cm}]





\tableofcontents

\section{Introduction}
\label{ch:intro}

The study of systems of strongly correlated electrons that display
non-Fermi-liquid behavior has attracted widespread interest in recent
years, fueled in part by their possible relevance to heavy-fermion
compounds \cite{Cox87,SML91,AT91} and high-$T_c$ superconductivity
materials \cite{CJJ89,EK92,GVR93}. On the theoretical front, 
one of the consequences was 
a renewed interest in various multi-channel Kondo models,
some of which were predicted by Nozi\`eres and Blandin \cite{NB80}
to contain non-Fermi-liquid physics.
Some of the most recent advances were made by
Affleck and Ludwig (AL) (see \cite{Lud94a} and
references therein),
who developed an exact conformal field theory
(CFT) solution for the $T=0$ fixed point of  the multichannel
Kondo models.
On the experimental front,
 an experiment performed by two of us (RB)
\cite{RB92,Ralph93},
that investigated certain zero-bias anomalies (ZBAs) in the conductance
of quenched  Copper nanoconstrictions,
has emerged as  a potential experimental realization of the
2-channel Kondo (2CK) model
and the corresponding non-Fermi-liquid physics
\cite{RLvDB94,HKH94,HKH95,JvDthesis}.
Although criticisms of the 2CK interpretation \cite{WAM95,MF96} and alternative
mechanisms for the ZBAs have been offered \cite{KK86,KRS95},
the 2CK scenario has recently received important additional suppport from
experimental results on ZBAs in constrictions made from
Titanium \cite{ULB96} and from metallic glasses \cite{KSvK95,KSvK96}.

In a series of three papers (I, II, III)
we shall present a detailed analysis of these ZBA experiments
and their 2CK interpretation.  The present paper (I) is  
a comprehensive review of the ZBA experiments
 that attempts to integrate all experimental results 
on the quenched Cu, Ti and metallic glass constrictions 
into a coherent picture (while postponing
all formal theoretical developments to papers II and III).
Paper II contains a calculation of the non-equilibrium conductance through a
nanoconstriction containing 2CK impurities, which is compared with
the Cu experiments.
Paper III, which is the only paper of the three
that requires knowledge of AL's conformal field theory solution
of the 2CK model, describes a bosonic reformulation \cite{ML95}
of their theory that is considerably simpler than those used previously
and is needed to derive certain key technical results used in paper II.

Let us begin by briefly summarizing the quenched Cu  ZBA experiments
and how they inspired the theoretical work presented 
in papers II and III.

\begin{figure} 
\vphantom{.} \vspace{2cm}
\centerline{\psfig{figure=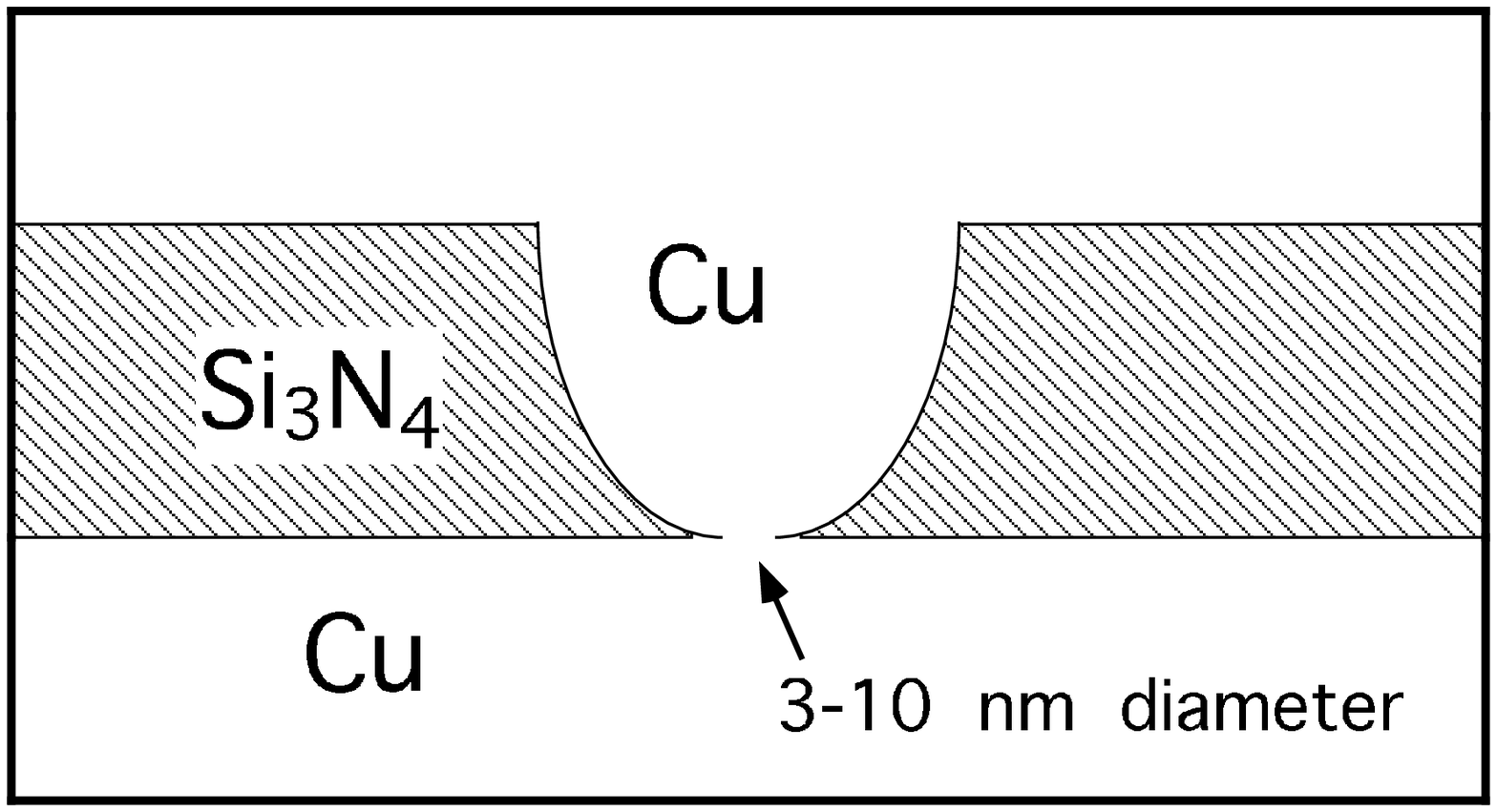,height=3cm}}
\vphantom{.} \vspace{2cm}
\caption[Cross-sectional schematic of a metal nanoconstriction.]{
\label{fig:nanoconstriction}
Cross-sectional schematic of a metal nanoconstriction.
The hole at the lower edge of the Si${}_3$N${}_4$ is so small that
this region completely dominates the resistance of the device.}
\end{figure}

RB used  lithographic techniques to
manufacture quenched Cu constrictions of diameters as small as 3~nm (see
Fig.~\ref{fig:nanoconstriction}),
and studied the conductance $G(V,T,H)$ through  the so-called nanoconstriction
(or point contact)
as a function of voltage $(V)$, temperature $(T)$ and magnetic field
$(H)$. Their constrictions were so small that they were able
to detect electron scattering at the level of individual impurities or
defects in the constriction.
Since the energy dependence of the scattering rate
can be extracted from the voltage dependence of the conductance,
such an experiment probes the
actual electron-impurity scattering mechanism.

For very small $eV/\kb $ and $T$ ($< 5 K$),
RB observed non-ohmic ZBAs in the
$V$- and $T$-dependence of the
conductance signals of unannealed, ballistic nanoconstrictions.
The qualitative features of these anomalies
(such as their behavior in a magnetic field, under annealing
and upon the addition of static impurities),
which are reviewed in detail in the present paper,
lead to the proposal  \cite{RB92} that the ZBAs 
are caused  by 
a special type of defect   in the nanoconstrictions, namely 
two-level systems (TLSs). 
This proposal has recently received strong support from a 
number of subsequent, related experiments 
(briefly reviewed in section~\ref{relatedexp})
by  Upadhyay {\em et al.}\/ 
on Ti constrictions  \cite{ULB96} and by Keijsers {\em et al.}\/
on metallic-glass constrictions   \cite{KSvK95,KSvK96}.


There are at least two theories for how TLSs can cause ZBAs in 
nanoconstrictions. In the first, based on Zawadowski's
{\em non-magnetic Kondo model}\/ \cite{Zaw80,VZ83}, the interaction
between TLSs and conduction electrons is described, at
sufficiently low energies, by the 2CK model
(reviewed in Appendix~\ref{ch:bk} of paper~II),
leading to an energy dependent scattering rate and hence a ZBA.
In the second, Kozub and Kulik's  theory of {\em TLS-population
spectroscopy}\/ \cite{KK86,KRS95}, the ZBA is attributed to a 
$V$-induced non-equilibrium occupation of the upper and 
lower energy states of the TLSs (see Appendix~\ref{KKpop}).

Though the two theories make quite similar predictions 
for the shape of the 
ZBA, they make different predictions for  the $V/T$-scaling 
behavior of $G(V,T)$. Whereas Kozub and Kulik's theory 
predicts that $G(V,T)$ does not obey any $V/T$-scaling relation
at all, the 2CK scenario predicts \cite{RLvDB94} that 
in the regime $T \ll \Tk $ and $eV \ll \kb \Tk$ (where $\Tk$
is the Kondo temperature), the conductance $G(V,T)$ should obey
a scaling relation of the form
\be
\label{scaans}
        {G(V,T) - G(0,T) \over T^{\alpha}} \, = \, F(eV / \kb T) \; ,
\ee
where $F(x)$ is a sample-dependent scaling function.
Moreover, AL's CFT solution of the 2CK problem suggested
that by scaling out non-universal constants, it should be possible
to extract from $F(x)$ a {\em universal}\/
(i.e. sample-independent) scaling curve $\Gamma (x)$,
 and that the conductance exponent
$\alpha$ should have the universal non-Fermi-liquid
value $\alpha = \half$, in striking contrast to the usual Fermi-liquid
value \cite{Noz74} of $\alpha = 2$.
Since no calculation
had been provided in Ref. \cite{RLvDB94} to support
the statement  that $\alpha = \toh$,
its status up to now has been that of an informed guess
rather than a definite prediction, a situation that is remedied
in papers II and III.

A detailed scaling analysis
\cite{RLvDB94} showed that the data of RB indeed
do obey the above scaling relation, with $\alpha = 0.5 \pm 0.05$.
It should be emphasized that the verification of scaling
was  a very significant experimental
result: firstly, the scaling relation~(\ref{scaans}), by combining
the $V$- and $T$-dependence of $G(V,T)$ for arbitrary ratios of $V/T$,
contains much more information than statements about the separate 
$V$- or $T$-dependence would; and secondly,
an   accurate experimental
determination of the scaling exponent $\alpha$ 
is possibly only by a  scaling analysis of all the data (for
a detailed review of this central ingredient of the data analysis,
see section~\ref{ch:scaling}). Accurate knowledge of $\alpha$
is very important, since $\alpha$ succinctly characterizes
the low-energy critical properties of the physics, enabling one
to eliminate many otherwise plausible candidate theories
for the ZBA (such as that by Kozub and Kulik).

The experimental value for $\alpha$ agrees remarkably well
with the CFT prediction of $\alpha = \half$;
furthermore, the scaling curve $\Gamma(x)$ is indeed the same for all three
samples studied in detail by RB, in accord with the CFT expectation
that it should be universal and hence sample-independent.
Thus, this result considerably strenghtened the case for 
the 2CK interpretation of the RB experiment,
within which the experimental demonstration that $\alpha = {1 \over 2}$
is, remarkably, equivalent to the direct observation of non-Fermi-liquid
physics.

Nevertheless, this scaling
behavior can conceivably also be accounted for by some other theory.
Indeed, Wingreen, Altshuler and Meir  \cite{WAM95}(a)
have pointed out that an exponent of $\alpha = \half$
also arises within an alternative interpretation of
the experiment, based not on 2CK physics but the physics of
disorder.
(We believe that this interpretation is in conflict with
other important experimental
facts, see section~\ref{p:WAMWAM}).

It is therefore desirable to develop additional quantitative criteria
for comparing experiment to the various theories.
One possible criterion 
is the  scaling function $\Gamma (x)$.
A very stringent quantitative test of any theory for
the RB experiment would therefore be to calculate the
universal scaling function $\Gamma (x)$,
which should be a fingerprint of the theory, and compare
it to experiment.
Papers II and III are devoted to this task:
 $\Gamma(x)$ is calculated analytically within the framework of
the 2CK model and its exact CFT solution by AL,
and the results are compared to the RB experiment.
When combined with recent numerical results
of Hettler {\em et al.} \cite{HKH94},
agreement with the experimental scaling curve is obtained,
thus lending further quantitative support to the 2CK interpretation
for the Cu constrictions.

The main conclusion of our work
is that the 2CK interpretation can qualitatively and quantitatively
account for all the scaling
properties of the conductance measured in the ZBAs of Cu point
contacts.  The Ti and metallic glass results
add further evidence in support of the 2CK 
interpretation, as opposed to other proposed mechanisms.
However, we shall note that the 2CK model does not account for
two phenomena observed in the quenched Cu samples.  Firstly, 
the magnetic field dependence of the
low-bias conductance is rather strong (the 2CK explanation
for the field dependence that was offered in Ref.~\cite{RLvDB94}
does not seem to survive closer scrutiny, as discussed in
Appendix~\ref{sec:magfield}).
Secondly, in many (but not all) Cu constrictions
the conductance undergoes very sudden
transitions at certain voltages $V_c$ \cite{Ralph93,RB94}
(see (Cu.9) of section~\ref{sec:data})
if $T$ and $H$ are sufficiently small.
These voltages can be  rather large
($V_c$ typically ranges between 5 and 20~mV),
implying that some new, large energy scale is involved.
These two phenomena are not generic to TLS-induced ZBAs, however, 
since they are observed neither in metallic-glass constrictions
nor in Ti constrictions, 
which in fact conform 
in all respects to what is expected for 2CK physics.
We shall suggest that the two phenomena involve (as yet poorly
understood) ``high-energy'' physics
associated with the strongly-interacting system of electrons and
atomic tunneling centers.
Such physics is beyond the scope of the existing 
2CK model and its CFT treatment, 
which deals only with the ``low-energy'' aspects of the problem.


Paper  I  is organized as follows:
In section~\ref{sec:exp.det} we describe the fabrication
and characterization of nanoconstrictions,
and summarize some elements of ballistic point
contact spectroscopy in section~\ref{sec:pcspectroscopy}.
In section~\ref{sec:data}
we summarize the main experimental facts associated with
the  ZBA in the Cu samples, which we state in the form
of nine properties, (Cu.1) to (Cu.9).
The 2CK interpretation
is presented in section~\ref{sec:TLSnano},
where its assumptions are summarized and critically discussed.
Section~\ref{ch:scaling} contains
a scaling analysis of the $G(V,T)$ data at $H=0$.
The related ZBA experiments on Ti and metallic-glass nanoconstrictions are
discussed in section~\ref{relatedexps}.
Finally, we summarize the results and conclusions of
this paper  in section~\ref{conclusions}.
Appendix~\ref{sec:elimination}  
describes experimental arguments for ruling out
a number of conceivable 
explanations for the ZBA that could come to mind as possible
alternatives to  the 2CK scenario. 
In Appendix~\ref{sec:magfield} we discuss
possible sources of magnetic field dependence in the 
2CK scenario, concluding it is essentially $H$-independent.

\section{The Nanoconstriction}
\label{sec:exp.det}

A schematic cross-sectional view of a typical nanoconstriction
(often also called a {\em point contact}\/)
is shown in Fig.~\ref{fig:nanoconstriction}. The
device is made in a sandwich structure. The middle layer is an
insulating Si${}_3$N${}_4$ membrane. This contains in one
spot a bowl-shaped hole, which just breaks through the lower edge
of the membrane to form a very narrow opening, as small as 3~nm
in diameter. This opening is so small  that the 
resistance signal,  measured between the top and bottom of 
the structure, is completely dominated by the  region within 
a few constriction radii of the opening.  Hence the resistance 
is sensitive to scattering from single defects in the constriction region.

To obtain the bowl-shaped hole in a Si${}_3$N${}_4$ membrane,
electron beam lithography and reactive ion etching are used in a technique 
developed by Ralls \cite{Rallsthesis}
(the details relevant to the present experiments
are described in Ref. \cite{Ralph93}, section 2.2). 
In ultra-high vacuum ($< 2 \times 10^{-10}$ torr)
and at room temperature the membrane is then rotated to expose
both sides while evaporating metal to fill the hole (thus forming
a  metallic channel through the constriction) and coat
 both sides of the membrane. A layer of at least 2000 $\AA$ of metal
(Cu or Ti in the work described below) 
is deposited on both sides of the membrane to form clean,
continuous films, and then  the devices are quenched (see property 
(Cu.1) in section~\ref{sec:data}).

\section{Ballistic Point Contact Spectroscopy}
\label{sec:pcspectroscopy}

A constriction is called {\em ballistic}\/ if   electrons
travel ballistically through it,
along semi-classical, straight-line
paths between collisions with defects or the walls of
the constriction.
This occurs if two conditions
are fulfilled: Firstly, it must be possible to neglect
effects due to the diffraction of electron waves,
i.e. one needs $1/k_{\sss F} \ll a$, where $a$~= constriction radius.
Secondly,  the constriction must be rather clean
(as opposed to disordered):
an electron should just scatter off impurities once or twice
while traversing the hole. One therefore needs $a \ll l$,
where $l$ is the electron mean free path.

The quenched Cu ZBA-devices of RB reasonably
 meet both conditions:  firstly, for Cu $1/k_{\sss F}
\simeq 0.1$nm, whereas $a$ is of order
2-8 nm [as determined from
the Sharvin formula for the conductance, \Eq{eq:2.sharvin}].
Secondly, for clean, annealed devices
$l \sim 200$~nm (as determined from the residual bulk resistivity).
For devices containing structural defects,
$l$ is reduced to about $l \gsim 30$~nm [see (Cu.4)],
which is still about twice the constriction diameter.
Thus,  we shall henceforth regard the quenched Cu ZBA-devices
as ballistic constrictions.

Some aspects of the theory of transport through ballistic
constrictions \cite{JvGW80,DJW89}
are reviewed in Appendix~\ref{sec:sharvin} of paper II. 
Here we merely summarize the main conclusions.

The differential conductance has the general form
\be
\label{2.diffcond}
G(V) \equiv \left| {d I(V) \over d V} \right| = G_o + \Delta G (V) \; .
\ee
The constant $G_o$, the so-called Sharvin conductance,
arises from electrons that travel ballistically through the
hole without scattering. Sharvin showed that for a round hole,
\be
\label{eq:2.sharvin}
        G_o = a^2 e^2 m \ve_{\sss F} / (2 \pi \hbar^3),
\ee
where $a$ is the radius of the
hole, and hence the measured value of $G_o$ can be used
to estimate the size of the constriction.

Any source of scattering in the constriction that backscatters
electrons and hence prevents them from ballistically traversing the hole
gives rise to a backscattering correction $\Delta G$. If the electron
scattering rate $\tau^{-1} (\ve)$ is energy-dependent, $\Delta G(V)$
is voltage dependent. In fact,
one of the most important characteristics of ballistic
nanoconstrictions
is that the energy dependence of
 $\tau^{-1} (\varepsilon )$
can be directly extracted from the voltage dependence of
$\Delta G (V)$, which implies that  ballistic
nanconstrictions can be used to do spectroscopy of
electron-defect scattering.

If, for example, the voltage is
large enough to excite phonons ($>5$mV for Cu), the $I$-$V$
curve is dominated by electron-phonon scattering. In this case,
it can be shown that at $T=0$,
$
        \Delta G(V)
        = -  \left(
        4 e^2 m^2 v_F a^3 / 3 \pi \hbar^4
        \right)
        \tau^{-1} (e V),
$
where $\tau^{-1} (\varepsilon') \equiv
\int_0^{\varepsilon'} d \varepsilon \,
        \alpha^2 F_p (\varepsilon) $
is the relaxation rate for an electron
at energy $\varepsilon'$ above the Fermi surface.
Thus, due to phonon-backscattering processes,
the conductance of any point contact drops markedly
at voltages large enough to excite phonons [$V > 5$~meV for
Cu, see Fig.~\ref{fig:dip+spike}(a)].
Furthermore, the function
$ \alpha^2 F_p (e V)$,
the so-called {\em point contact phonon spectrum}\/,
can be directly obtained from
$\partial_V  \Delta G(V)$.
For any clean, ballistic Cu nanoconstriction, 
$\partial_V \Delta G(V)$ should give
the same function $\alpha^2 F_p(eV)$,
characteristic of the phonon spectrum,
and indeed nanoconstriction
 measurements thereof agree with other determinations of
$\alpha^2 F_p$. However, the amplitude of the phonon-induced
peaks is reduced dramatically if there is significant
elastic scattering due to defects or impurities in
the constriction region, as has been modeled theoretically
\cite{YS86} and demonstrated experimentally \cite{LYSN80}.
Therefore, \label{sec:phonons}
comparing \label{p:phononpeak}the point contact
phonon spectrum  of a given  point contact
 to the reference spectrum of a clean point contact provides an
important and reliable tool for determining
whether the point contact is clean or not.

\label{sec:defectscattering}

For voltages below the  phonon threshold ($V<5 $mV for Cu),
the $V$-dependence of $\Delta G(V)$
is due to scattering off defects. For a
set of defects at positions $\vec R_i$,
with an isotropic, elastic, but energy-dependent scattering
rate $\tau^{-1} (\ve)$,
the backscattering conductance has the form \cite{JvDthesis}
\bea
\label{eq:2.diffconddef}
\lefteqn{\Delta G (V)
        =  - (\tau(0) e^2/h )
        \!\! \int^\infty_{-\infty}  \!\! d \omega
        [- \partial_{\omega} f_o
        (\hbar \omega) ] } 
\\ && \nonumber
\times \sum_i b_i
        \half  \Bigl[ \tau^{-1} (\hbar \omega - \half e V a^+_i) +
        \tau^{-1} (\hbar \omega -   \half e V a^-_i) \Bigr] \; .
\eea
We factorized out the constant $ \tau(0) e^2/h$ to ensure that
$\Delta G$ has the correct dimensions and order of magnitude.
We assume that the resistance contribution from each defect
may be calculated independently -- that is, that quantum
interference for electrons scattering from multiple defects
may be ignored. 
The $a_i$ and  $b_i$ are (unknown) constants  of order
unity that characterize all those details of scattering
by the $i$-th impurity that are energy-independent and
 of a sample-specific, geometrical nature.
The $b_i$ account for the fact that the probability that an electron
will or will not traverse the hole after being scattered off the
$i$-th impurity depends on the position of the impurity relative
to the hole. The $a_i$ account for the fact that  impurities that
are at different positions $\vec R_i$ in the nanoconstriction feel
different effective voltages (because the amount by which the
non-equilibrium electron distribution function at $\vec R_i$
differs from the equilibrium Fermi function $f_o$ depends on $\vec R_i$).

In spite of the presence of the many unknown constants
$a_i$, $b_i$, we shall see that it is nevertheless possible to extract
general properties of $\tau^{-1}(\ve)$ from the measured
$\Delta G(V,T)$ data. For example, from \Eq{eq:2.diffconddef}
one can deduce  that if
\bea
\label{eq:genscaling}
\lefteqn{       \tau^{-1} (\ve, T) - \tau^{-1} (0,T) \propto
        \left\{
                \begin{array}{c}
                        \ln \mbox{[max$(T, \ve)$]}  \, ,
                \\
                        T^\alpha \tilde \Gamma (\ve/T) \, ,
                \end{array}
        \right.}
\\ && \nonumber
\quad \mbox{then} \quad
        \Delta G(V,T) \propto
        \left\{
                \begin{array}{c}
                  \ln \mbox{[max$(T, V)$]} \,  ,
                \\
                        T^\alpha F(V/T) \; ,
                \end{array}
        \right.  \;
\eea
where $\tilde \Gamma $ and $F$ are scaling functions.

\section{Experimental Facts for Quenched Cu Samples }
\label{sec:data}

In this section we summarize  the experimental facts
relevant to the ZBA in quenched Cu samples.
Our interpretation of these facts is postponed to later sections,
where some of them will be elaborated upon more fully,
and where most of the figures quoted below can be found.

\begin{figure}
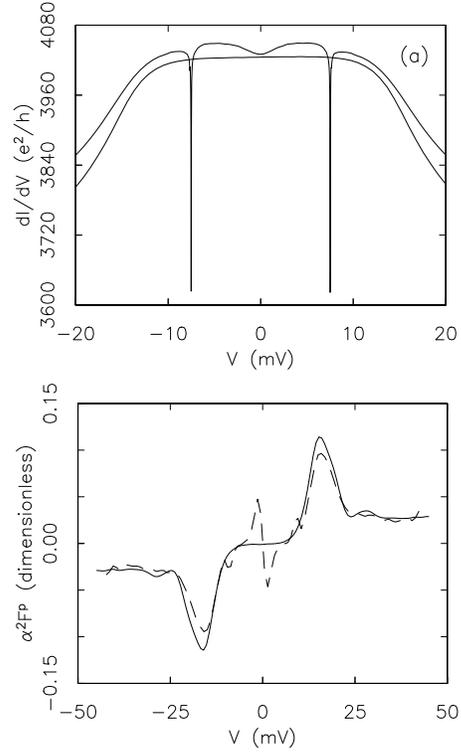

\vphantom{.}
\vspace{1.5cm}
\centerline{\psfig{figure=janfig2a.ps,height=6cm}}
\vspace{-1cm}
\centerline{\psfig{figure=janfig2b.ps,height=6cm}}
\caption[Typical conductance
curve for a constriction containing structural defects.]{
\label{fig:dip+spike}A typical conductance
curve for a constriction containing structural defects:
(a) The upper curve, showing a dip in conductance at $V=0$ and
voltage-symmetric spikes, is the differential conductance for an
unannealed Cu sample at 4.2 K. The lower curve shows the conductance
of the same device at 4.2 K, after annealing at room temperature for
2 days. The curves are not artificially offset; annealing changes
the overall conductance of the device by less than 0.5\%. (b)
Point contact phonon spectrum  at 2 K for the device before anneal
(dashed line) and after anneal (solid line).}
\end{figure}

The phenomenon to be studied is illustrated
by the upper differential conductance curve in Fig.~\ref{fig:dip+spike}.
Its three essential features are the following:
Firstly,  the differential conductance shows a drop
for $|V| > 5 $~mV, due to the excitation of phonons,
a process which is well understood (see
section~\ref{sec:pcspectroscopy}).
Secondly, there are
sharp voltage-symmetric conductance spikes at somewhat larger
voltages ($V_c$), 
called {\em conductance transitions} in Ref. \cite{RB92,RB95},
because in the DC conductance they show up
as downward steps with increasing $V$
(see figure~\ref{fig:condtrans} below).
Some of their complex properties are listed in point (Cu.9)
below.

Thirdly, the conductance has
a   voltage-symmetric dip near $V=0$;
this is  the so-called
{\em zero-bias anomaly}\/ (ZBA).
 As a sample is cooled, the temperature
at which the zero-bias features become measurable varies
from sample to sample, ranging from 10~K to 100~mK.
This paper is concerned mainly with the regime $V< 5 $~mV 
dominated by this ZBA.

The ZBA is a very robust phenomenon.
For decades it has been observed, though not carefully
investigated, in mechanical ``spear and
anvil'' point contacts made from a variety of
materials, see e.g. \cite{JMW78}. Even the
dramatic conductance transitions
have probably been  seen in early ZBA experiments
\cite{JvGW80},
though their presence had not been emphasized
there.\footnote{For example, Fig. 3C of \protect\cite{JvGW80}\protect\
shows a $d^2 I/d V^2 $ spectrum with sharp signals,
more or less symmetric about zero, that are consistent with
being derivatives of spikes in the $dI/dV$ conductance curve.
Note that these signals are too sharp to be spectroscopic signals
smeared by $kT$, but are indicative of abrupt transitions.}

\begin{figure}
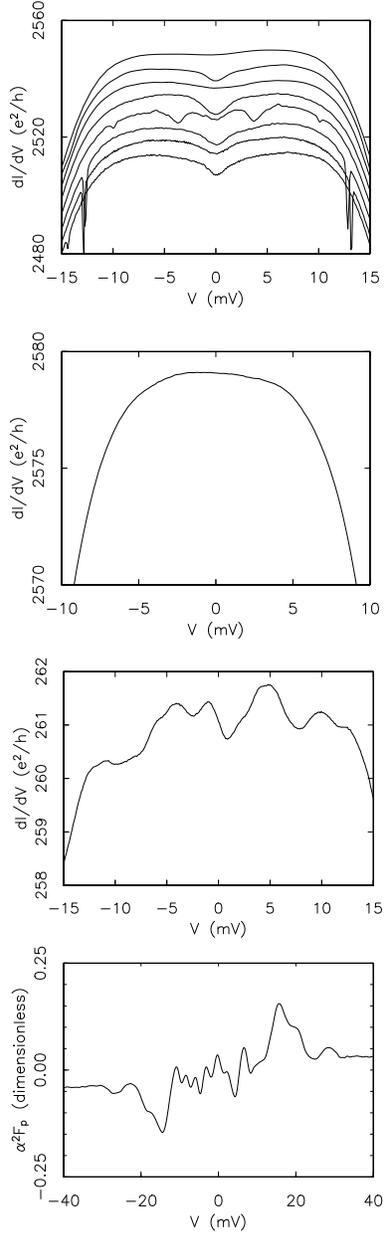
 
\vphantom{.}
\vspace{0.3cm}
\centerline{\psfig{figure=janfig3a.ps,height=5cm}}
\hspace{2.5cm}
\centerline{\psfig{figure=janfig3b.ps,height=5cm}}
\centerline{\psfig{figure=janfig3cd.ps,height=5cm}}
\vphantom{.}
\vspace{0.3cm}
\caption[Effect on the differential conductance
of (a) repeated thermal cycling,
(b) adding static disorder, (c) electromigration]{
\label{fig:thermalcycling}
(a) Differential conductance versus
voltage at 4.2~K for a Cu sample which underwent repeated thermal
cycling \protect\cite{Ralph93}\protect.
 The time sequence runs from the bottom curve to the top.
Curves are artificially offset. The first 2 excursions were to 77~K,
the next 5 to room temperature.
(b) 
 Differential conductance
for a Cu sample intentionally doped with 6 \% Au. Static impurities
reduce the electronic mean free path but completely
eliminate the zero-bias anomaly of interest to us.
\label{fig:staticdisorder}
(c)  
Differential conductance and
(d) point contact
spectrum  for a  Cu device at 1.8~K
in which disorder has been created by electromigration
(which means that a high bias
(100-500~mV) has been applied at low temperatures so that
Cu atoms moved around).
\label{fig:condfluctuations}}
\end{figure}

The advent of the mechanically very stable
nanoconstrictions employed by RB allowed a detailed
systematic study of the ZBA. Their findings
 are discussed at length in \cite{Ralph93} and \cite{RB95}.
We summarize them in the form of 9 important properties
of the ZBA in quenched Cu nanoconstrictions:

\begin{enumerate}
\item[(Cu.1)]\label{P1} {\em Quenching:}\/
ZBAs and conductance transitions [\fig{fig:dip+spike}(a)]
are found only in {\em quenched} Cu samples,
i.e.\ samples that are cooled to cryogenic
temperatures within hours after being formed by evaporation.
They are found in about 50\% of such samples, and in a variety of
materials, such as Cu, Al, Ag and Pt. Cu was used in the samples
discussed below).
\item[(Cu.2)] {\em Amplitude:}\/ 
Typical values for $G(V \! = \! 0)$ vary from 2000 to 4000~$e^2/h$.
The anomaly
is only a small feature on a very big background conductance:
its  amplitude [$ G_{max} - G(V \! = \! 0)$]
varies from sample to sample, from a fraction of $e^2/h$ to
as large as $70 e^2 / h$ at 100 mK. It's sign is always
the same, with $G(V,T)$ increasing from
$G(0, T_o)$ as $V$ or $T$ are increased.
The sample (\#1 in Fig.~\ref{fig:6.7})
 showing best scaling (see (Cu.6) below)
had a maximum ZBA amplitude of about $20 e^2/h$.

\item[(Cu.3)] {\em Annealing:}\/
\\
(a) After annealing at room temperature for several days,
the ZBA and conductance spikes disappear,
and the conductance curve looks like that of a completely
clean point contact [see lower curve in Fig.~\ref{fig:dip+spike}(a)].
\\
(b) Nevertheless, such annealing changes the total conductance by not
more than 1\% or 2\% (both increases and decreases have been observed),
indicating that the overall structure of the constriction
does not undergo drastic changes.
\\
(c) Upon thermal cycling, i.e. brief (several minutes)
excursions to room temperature and back,
the amplitude of the ZBA and the position $V_c$ of the
conductance transitions change dramatically and non-monotonically
[see Fig.~\ref{fig:thermalcycling}(a)].
The complexity of this behavior suggests that
the thermal cycling is causing changes in the position of
defects within the constriction, and that the ZBA is very
sensitive to the precise configuration of the defects.

\item[(Cu.4)] {\em Effect of disorder:}\/
\\
(a) \label{p:P3} If static disorder is intentionally introduced
into a nano\-constriction by adding 1\% or more of impurity
atoms such as Au to the Cu during evaporation,
the zero-bias conductance dip and conductance spikes {\em disappear
completely}\/ [see Fig.~\ref{fig:staticdisorder}(b)].
Likewise, the signals are absent in samples for which
water is adsorbed onto the Si${}_3$N${}_4$ surface before metal
deposition (the standard sample fabrication procedure
therefore involves heating the sample to $\sim 100^\circ$C in vacuum,
or exposing it for several hours to ultraviolet light in vacuum,
before the final metal evaporation is done).
\\
(b) When a strongly disordered region is created near
the constriction (by electromigration: a high bias
(100-500~mV) is applied at low temperatures so that
Cu atoms are moved around, a method controllably demonstrated
in \cite{Rallsthesis,RB88,RRB89}),
the conductance shows no ZBA either, but instead
small-amplitude, voltage-dependent (but aperiodic)
conductance fluctuations at low voltage
[see Fig.~\ref{fig:condfluctuations}(c), (d)]. 
That these are characteristic of
strongly disordered constrictions and can be
interpreted as universal conductance fluctuations due to
quantum interference,  was established in a
separate investigation
\cite{Holweg91}, \cite[chapter~4]{Ralph93}, \cite{RRB93}.
\\

\item[(Cu.5)] {\em Phonon spectrum:}\/
\label{P5}
For quenched samples, in the  point contact phonon spectrum
the longitudinal phonon peak near 28 mV is not well-defined,
and the total amplitude of the  spectrum
is  smaller by about 15\% than after annealing.
After annealing,  the  longitudinal phonon peak reappears
and the spectrum
corresponds to that of clean ballistic point contacts.
Both  these differences indicate (see p.~\pageref{p:phononpeak})
that the elastic mean free path $l$
in the annealed samples is somewhat longer than in the quenched samples.
From the phonon spectrum of the latter,
$l$ can be estimated  (see section~\ref{sec:exp.det})
to be $l \gsim 30$~nm [for the sample shown in Fig.~\ref{fig:dip+spike}(a)],
still more than about twice the constriction diameter
for that device.
Note also that the point contact phonon spectrum for a quenched device
[Fig. \ref{fig:dip+spike}(b)]
is qualitatively very different from that of a strongly disordered
constriction [Fig. \ref{fig:condfluctuations}(d)].
These facts, viewed in conjunction with (Cu.3b) and (Cu.4c),
imply that the Cu constrictions displaying ZBAs are still
rather clean and ballistic.\/
\item[(Cu.6)]
{\em $V/T$ scaling}\/ (to be established in detail in 
section~\ref{ch:scaling}):
\\
(a) At $H=0$, the
conductance obeys the following scaling relation
if  both $ V < \Vk$ and $T< \Tk$,
but for arbitrary ratio $v = e V/\kb T$:
\be
\label{eq:2.scalingform}
       {G(V,T) - G(0,T) \over  T^{\alpha}} =  F (v) \; .
\ee
Relation~(\ref{eq:2.scalingform}) allows a large number of data curves
to be collapsed onto a single, sample-dependent scaling curve
[e.g.\ see Figs.~\ref{fig:6.10}(a) and \ref{fig:6.11}(b) below].
The departure of individual curves from the low-$T$ scaling curve in
Figs.~\ref{fig:6.10}(a) and \ref{fig:6.11}(b) indicates that $V$ or $T$
has surpassed the crossover scales $\Vk$ or $\Tk$.
From the data, these are related roughly by
$e \Vk = 2 \kb \Tk$, with $\Tk$ in the range 3 to 5~K.
\\
(b) $F(v)$ is a sample-dependent scaling function with the
properties $F(0) \neq 0$ and $F (v) \propto v^{\alpha}$ as
 $v \to \infty$, and the scaling exponent is found to have
the value $\alpha = 0.5 \pm 0.05$.
\\
(c)
By scaling out sample-dependent constants,
it is possible to extract from $ F(v)$ a
``universal'' scaling function $\Gamma(v)$ [shown in
Fig.~\ref{fig:6.16}(b) below]. $\Gamma(v)$
 is universal in the sense that it is
indistinguishable
for all three devices  for which a  scaling analysis was
carried out (they are called sample 1,2 and 3 below).
\item[(Cu.7)]
{\em Logarithms:}\/
For $V$ or $T$ beyond the cross-over scales $\Vk$ or $\Tk$,
$G(V,T)$ deviates markedly from the scaling behavior of (Cu.6)
and behaves roughly logarithmically:
For $H=0$ and fixed, small $T$, the  conductance
goes like $\ln V$ for $V > \Vk$ [Fig.~\ref{fig:Kondologs}(a)];
similarly, for $H=V=0$ and $T > \Tk$, the conductance
goes like $\ln T$ [Fig.~\ref{fig:Kondologs}(b)].
\item[(Cu.8)]
{\em Magnetic field:}\/
\\
(a) When a magnetic field (of up to 6~T)
is applied, the amplitude of the ZBA in Cu devices decreases
[see Fig.~\ref{fig:magnetic}(b)]. The change in amplitude
can be as large as 24 $e^2/h$ if $H$ changes from 0 to 6~T.
For sufficiently small $H$ ($<1$T), at fixed $T$ and
$V=0$,  the magnetoconductance roughly follows
$G(H,T) \propto |H|$ (see Fig.~\ref{fig:6.17} below).
However,  the available data is insufficient
to establish linear behavior beyond doubt,
and,  for example, would also be compatible with a
 $|H|^{1/2}$-dependence.
\\
(b) The ZBA dip undergoes no
Zeeman-splitting in $H$, in constrast to
the Zeeman splitting that {\em is}\/ found for devices intentionally
doped with magnetic impurities such as Mn [see
Fig.~\ref{fig:magnetic}(a)].
\item[(Cu.9)]\label{P9} {\em Conductance transitions:}\/
\\
(a) Voltage-symmetric conductance transitions (spikes in the differential
conductance at certain ``transition voltages'' $V_c$,
see Fig.~\ref{fig:dip+spike})  occur only in quenched  point contacts
that show ZBAs, but occur in at least 80\% of these.
The spikes disappear under annealing, just as the ZBA does (Cu.3a).
\\
(b) (i) A single sample can show several
such conductance transitions (up to 6 different $V_c$s
have been observed in a single sample). (ii) If
 $T$ and $H$ are small (say $T\lsim 1$K, $H \lsim 0.5$T),
$V_c$ is typically rather large, with typical values
ranging between 5 and 20 mV, well above the typical
voltages associated with the ZBA (i.e. $V_c > \Vk$).
The spikes have a very complex behavior as a
function of temperature ($T$) and magnetic field ($H$),
including (iii) a hysteretic $V$-dependence, (iv) a bifurcation
of single spikes into two separate ones $(V_{c1}, V_{c2})$ when $B \neq
0$ (Fig.~\ref{fig:Vcmotion}), (v)
the $H$-dependent motion of the  spike positions $V_c (H) \to 0$
when $H$ becomes sufficiently large (Figs.~\ref{fig:Vcmotion},
\ref{fig:condtrans}), and (vi) a very rapid
narrowing of the peaks with decreasing $T$. They are
 described  at length, from a phenomenological point
of view, in Ref. \cite{RB95}.
\end{enumerate}

Any theory that purports  to explain the ZBA in Cu constrictions
must be consistent with  all of the above experimental facts.
An extention of this list to include the results of
the recent related ZBA experiments 
by Upadhyay  {\em et al.}\/ on Ti constrictions and 
 by Keijsers {\em et al.}\/
\cite{KSvK96} 
on metallic-glass constrictions is presented in section~\ref{relatedexps}.

In the next section, we shall argue that the 2CK scenario
provides the most plausible
interpretation of the above experimental facts.
A  number of alternative explanations for the ZBA
that could come to mind  are discussed in 
Appendix~\ref{sec:elimination}, but all are found
to be in conflict with some of the above facts.

\section{The 2-channel Kondo (2CK) Interpretation}
\label{sec:TLSnano}

In this section, we develop  the 2CK interpretation of the
ZBAs in quenched Cu constrictions.
{\em It attributes the ZBA to the presence in the constriction region
of dynamical structural defects, namely TLSs,
 that interact with conduction electrons
according to the non-magnetic Kondo model,  which renormalizes at low energies
to the \nflr\ of the 2CK model.}\/ 
We begin by briefly recalling in section~\ref{sec:TLSnano}.A
some properties of two-level systems
(or slightly more generally, dynamical two-state systems) in metals.
Successes and open questions of the 2CK scenario
are discussed in subsections~\ref{sec:TLSnano}.B and
\ref{sec:TLSnano}.C, respectively, 
and its key assumptions are listed, in the form of a summary,
in subsection~\ref{sec:TLSnano}.D.

\subsection[Two-state systems]{Two-State Systems}
\label{sec:twostate}

A dynamical
two-state system (TSS), is an atom or
group of atoms that can move between two different
positions inside a material \cite{AHV72}.
In the absence of interactions,
its behavior is governed by a double-well potential,
generically depicted in Fig.~\ref{fig:doublewell},
with asymmetry energy $\Delta_z$,
tunneling matrix element $\Delta_x$. The corresponding Hamiltonian is
\be
\label{HTLSonly}
        H_{TSS} =
         \half \left( \Delta_z \tau^z  + \Delta_x \tau^x \right) \;,
\ee
where $\tau^x$ and $\tau^z$ are Pauli matrices acting in
the two-by-two Hilbert space spanned by the states $|L\rangle$
and $|R \rangle$, describing  the fluctuator in the left or right well.

Depending on the parameters of the potential, the
atom's motion between the potential wells is
classified as either slow, fast or ultrafast,
with hopping rates $\tau^{-1} < 10^8 s^{-1}$,
$10^8 s^{-1} < \tau^{-1} < 10^{12} s^{-1}$
or $\tau^{-1} > 10^{12} s^{-1}$, respectively \cite{CZ95}.
{\em Slow}\/ two-state systems, called {\em two-state fluctuators},
have large barriers and neglibibly small $\Delta_x$, and the
motion between wells occurs due to thermally activated hopping
or incoherent quantum tunneling.
{\em Fast}\/ two-state systems have sufficiently
small barriers and sufficiently large $\Delta_x$
that {\em coherent tunneling}\/ takes place
back and forth between the wells.
 Such a system is known as a
{\em two-level tunneling system}\/ (TLS),
because its physics is usually dominated by
its lowest two eigentstates (even and odd
linear combinations of the lowest-lying eigenstates
of each separate well), whose
eigenenergies differ by $\Delta = (\Delta_z^2 + \Delta_x^2)^{1/2}$.
{\em Ultra-fast}\/ two-state systems have such a large $\Delta_x$
 that $\Delta$ too becomes very large, so  that at low
temperatures only the lowest level governs the physics.

\subsubsection{Slow Fluctuators}
\label{slowfluctuators}

The fact that two-state systems
in metal nanoconstrictions can
influence the conductance was demonstrated
by Ralls and Buhrman \cite{RB88,RRB89,RRB93},
who observed  so-called telegraph signals 
in {\em well}\/-annealed devices
(at rather high temperatures of 20-150K). These are 
slow, time-resolved fluctuations
(fluctuation rates of about $10^3 s^{-1}$)  of the conductance
between two (or sometimes several)
discrete values, differing by fractions of $e^2/ h$, which
can be attributed to the fluctuations of a 
{\em slow}\/
two-state fluctuator in the constriction region.

Such telegraph signals 
were also observed by Zimmerman {\em et al.}\/ \cite{ZGH91,GZC92},
\label{p:bismuth} who studied the conductance
of polychrystalline Bi films, a
highly disordered material with presumably large
numbers of two-state systems.
They were able  to measure the  parameters
of individual slow fluctuators
directly, finding values for the asymmetry energy $\Delta_z$
ranging from as little as $0.08$~K to about 1K.
They also demonstrated that in a disordered environment
the asymmetry energy of a TLS is a random, non-monotonic
function  of the magnetic field, $\Delta_z = \Delta_z (H)$
(as predicted earlier in Ref. \cite{AS89}), and
hence can be ``tuned'' at will by changing $H$.
The reason is, roughly, that $\Delta_z$ depends on the difference
$\delta \rho = \rho_L- \rho_R$ between
the local electron densities at the two
minima of the TLS potential. Due to quantum interference
effects that are amplified by the presence of disorder,
changes in $H$ can induce random changes in 
$\delta \rho$ and hence also in $\Delta_z$.

Unfortunatly,   experiments on  {\em slow}\/ fluctuators do not yield
any direct information on the parameters to be expected for
fast ones, since their parameters fall in different ranges.

\subsubsection{Two-Level Systems}
\label{non-mag}

Fast fluctuators or TLSs
presumably have the same microscopic nature and origin
as slow fluctuators, being composed
of atoms or small groups of atoms which move between two metastable
configurations, but with much lower barriers.
Therefore,  they anneal away quicker than slow fluctuators,
which is why they were not seen in the above-mentioned Ralls-Buhrman
experiments on well-annealed samples  \cite[p.~265]{Ralph93}.
Also, whereas  slow fluctuators ``freeze out'' as
$T$ is lowered (which is why they don't play a role
in the ZBA regime of $T< 5$K), at low $T$
fast fluctuators continue to undergo
transitions by tunneling quantum-mechanically between the wells.

A fast fluctuator or TLS interacting with conduction
electrons is usually described by 
the {\em non-magnetic}\/ or
{\em orbital Kondo model,}\/
studied in great detail by Zawadowski and coworkers  \cite{Zaw80,VZ83}
(it is defined and reviewed
in more detail in Appendices~\ref{sec:bk} and \ref{app:bk} of paper II;
for other reviews, see \cite{CZ95,ZV92,Zarand97}):
\bea
\label{HTLSall} \label{HTLS}
        H &=& H_{TSS} + \sum_{\vec k} \ve_{\vec k}
         c^{\dagger}_{\vec k \sigma} c_{\vec{k}' \sigma}
\\
       && \; + \;
\nonumber
        \sum_{\vec k \vec{k}'}
        c^{\dagger}_{\vec k \sigma}
        \left[ V^0_{\vec k \vec{k}'}
        +  V^x_{\vec k \vec{k}'} \tau^x
        +  V^z_{\vec k \vec{k}'} \tau^z
        \right]  c_{\vec{k}' \sigma} \; .
\eea
Here $c^{\dagger}_{\vec k \sigma}$ creates an electron with momentum
$\vec k$ and Pauli spin $\sigma$.
The terms $V^o $ and  $V^z \tau^z$ describe
{\em diagonal}\/ scattering events in which the TLS-atoms do not tunnel
between wells. The term $V^x \tau^x$ 
describes so-called
{\em electron-assisted tunneling}\/ processes. During these,
electron scattering does lead to tunneling, and hence
the associated bare
matrix elements are much smaller than
for diagonal scattering: $V^x / V^z \simeq 10^{-3}$.

Zawadowski and coworkers
 showed  that the electron-assisted term $V^x \tau^x$
 renormalizes to substantially larger values as the temperature is lowered
 (as does a similar $V^y \tau^y$ term that is generated under
renormalization). At sufficiently
low temperatures (where $V^z \simeq V^x \simeq V^y$),
the non-magnetic Kondo model was shown \cite{MG86} to be equivalent
to the standard 2-channel Kondo (2CK) model, 
with an  effective interaction of the form
\be
\label{Ibk.HeffV}
        H^{eff}_{int} =
         v_{\sss K}  \int \! \! d \ve \! \!  \int \! \! d \ve' \!
        \sum_{\alpha, \alpha' } \sum_{\sigma \sigma'}
        c^{\dagger}_{\ve \alpha \sigma} 
        \, 
        \left( \half  \vec \sigma_{\alpha \alpha'}
          \, \cdot \half \vec \tau \right) \,
        c_{\ve \alpha' \sigma}    \; .
 \ee
The two positions of the fast fluctuator in the $L$- and $R$ wells
correspond to the spin up and down of a
magnetic impurity (and $L$-$R$ transitions to
impurity spin flips). The electrons are labelled by
an energy index $\ve$, a so-called pseudo-spin index $\alpha = 1,2$
(corresponding to those two combination of
angular momentum states about the impurity that
couple most strongly to the TLS), and the
Pauli spin index $\sigma = \uparrow, \downarrow$.
Evidently,
$\alpha$ plays the role of the electron's magnetic spin index
in the magnetic 2CK model, and 
since the effective interaction is diagonal in
$\sigma$ (which has two values),
$\sigma$  is the  channel index.

This (non-magnetic) 2CK model, with strong analogies to
the magnetic one, yields an 
electron scattering rate $\tau^{-1} (\ve,T)$ with  the properties
\cite{VZ83,AL93}
\bea
\label{2CKrates}
\lefteqn{        \tau^{-1} (\ve, T) - \tau^{-1} (0,T) \; \propto}
\\ && \nonumber 
       \qquad  \left\{
                \begin{array}{cl}
                        \ln \mbox{[max $(T, \ve)$]}  
        \qquad  & \mbox{if} \quad T > \Tk \; ,
                \\
                        T^{1/2} \tilde \Gamma (\ve/T) 
        \qquad  & \mbox{if} \quad \Delta^2/ \Tk < T \ll \Tk \; .
                \end{array}
        \right.
\eea
(The condition  $\Delta^2/ \Tk < T$
is explained in section~\ref{sec:Delta}.)
Hence, for $T> \Tk$ or $\Delta^2/ \Tk < T \ll \Tk$,  it 
yields [via \Eq{eq:genscaling}] a contribution to the 
conductance of $\delta \sigma (T)
\propto \ln T$ or $T^{1/2}$, respectively.
The latter is typical for the complicated
non-Fermi-liquid physics characteristic of the 2CK model
in the $T \ll \Tk$ regime.  In this respect the 
non-magnetic 2CK model differs 
in an important way from the (1-channel) magnetic
Kondo model, for which the low-$T$ scaling is of the
Fermi liquid form ($\propto T^2$).

\subsection[Successes of the 2CK Interpretation]{Successes 
of the 2CK Interpretation}
\label{successes}

We now turn to an interpretation of facts (Cu.1) to (Cu.9) in terms
of the 2CK scenario  \cite{RB92,RLvDB94}. Our aim here
is to sketch the physical picture underlying the scenario.
Those aspects that require detailed analysis, such
as the scaling behavior (Cu.6) and magnetic field
dependence (Cu.8), will be discussed more fully in subsequent sections.

\label{sec:strucdef}
{\em Qualitative features:}\/
The cooling and annealing properties (Cu.1) and (Cu.3) suggest that the ZBAs
are due to {\em structural }\/
defects or disorder that can anneal
away at high temperatures [although the well-resolved
phonon spectrum implies that 
only a small amount of such disorder can be present (Cu.5)].
This conclusion is reinforced by the remarkably complex and
non-monotonic behavior of the ZBA under thermal cycling (Cu.3c),
which indicates that the ZBA probes the
detailed configuration of individual defects,
not just the average behavior of the entire constriction region.
Subsequent experiments with Ti constrictions have shown
that the structural disorder is located in the ``bulk'' of the bowl-shaped
hole, not on its surface, and that 
it is caused by geometry-induced
stress occuring in the metal in the bowl-shaped part of the
constriction [see (Ti.1d), section~\ref{sec:stressed}].

By assuming that the ZBA is due to fast TLSs, i.e.\
a specific type of structural defect, the 2CK scenario accounts 
for all of the properties mentioned in the previous paragraph. Property (Cu.4a),
the disappearance of the ZBA upon the addition of 1\% Au atoms, 
can then be attributed to the TLSs being pinned by the additional
static impurities. 

{\em Logarithms and Scaling:}\/ 
Next, we assume that the TLS-electron interaction
is governed by Zawadowski's non-magnetic Kondo model,
which renormalizes to the 2CK model at low energy scales. This 
explains a number of further facts. Firstly, 
the non-magnetic nature of the interaction 
explains the absence of a Zeeman splitting in a
magnetic field (Cu.8b).
Furthermore, the fact that the 2CK scattering rate 
$\tau^{-1}(\ve, T)$ has a logarithmic 
form for $\ve > \Tk (>T)$ or $T > \Tk (>\ve)$   [see \Eq{2CKrates}]
accounts, via \Eq{eq:genscaling}, for
the asymptotic logarithmic $V$- and $T$ dependence (Cu.7)
of $G(V,T)$ for $V > \Vk (>T)$ or $T > \Tk ( > V)$.
Thus, we identify the experimental crossover 
temperature $\Tk$ ($\simeq 3$ to 
5K) of (Cu.6a)  with the Kondo temperature of the 2CK model.

Similarly, the scaling form of $\tau^{-1}(\ve, T)$ 
for $\ve, T \ll \Tk$ [see \Eq{2CKrates}]
accounts, via \Eq{eq:genscaling}, for the observed scaling
behavior (Cu.6) of  $G(V,T)$ for $V < \Vk$ and $T < \Tk$.
To be more particular, the very 
occurence of  scaling behavior (Cu.6a), 
and the fact that the experimental scaling curve 
$\Gamma (v)$ of  (Cu.6c) is universal, can be explained 
(see  section~\ref{sec:scalingansatz}) by assuming that 
the system is in the neighborhood of some fixed point.
Assuming this to be the 2CK \nfl\ fixed point, 
the experimentally observed scaling regime
can be associated with the theoretical expected scaling regime of 
$\Delta^2/ \Tk < T <\Tk$ and $V <  \Tk$.
Moreover, the \nfl\ value of  $\alpha = \half$ 
that is then expected for the scaling exponent
 (see section~\ref{sec:scalingansatz}) agrees precisely 
with the value observed for $\alpha$. Thus, 
within the 2CK interpretation, the experimental 
demonstration of $\alpha = \half$ is equivalent to
the direct observation of non-Fermi-liquid behavior.
Finally, it will be shown in paper II that
the shape of the universal scaling
curve $\Gamma (v)$ is also in quantitative agreement with the
2CK model.

{\em Number of TLSs:}\/ 
Each 2CK impurity in the constriction
can change the conductance by at most $2 e^2/h$.\footnote{\label{2e2h}
To see this, we note that in the unitarity limit
the scattering rate of electrons off a $k$-channel Kondo
defect is proportional to $k \sin^2 \delta$ (see e.g.\
\cite[Eq.(2.20)]{Hew93}), and the phase shift at the
intermediate-coupling fixed point is $\delta = \pi/ 2k$
\cite{VZZ88}. Thus, in the unitarity limit,
the contribution to the resistance  of a $k=2$ Kondo impurity
is the same as for $k=1$, namely $2 e^2/h$ (the 2 comes from Pauli spin).}
Therefore, the sample with the largest ZBA of $ 70 e^2/h$ 
(sample \#2 in Fig.~\ref{fig:6.7}) would require up to about 40 such 
TLSs in the constriction. However, this is still only a
relatively small amount of disorder (corresponding
to a density of about $10^{-4}$ TLSs per atom \cite[p.277]{Ralph93}\footnote{
For example, the $6.4 \Omega$ constriction
studied in \cite{RB92}
has a diameter of $\sim 13$~nm [estimated via the Sharvin formula
Eq.~(\protect\ref{eq:2.sharvin})], 
and there are $10^5$~Cu atoms within a sphere of
this diameter about the constriction. Assuming on the order of
$\sim$40 active TLSs, 
their density is therefore roughly of order $10^{-4}$/atom.
Although the constriction is believed to be crystalline,
not glassy, it is worth noting that this density of
TLSs is about the same as estimates for the total density of TLSs in
glassy systems.}).
The sample that showed the best scaling 
(sample \#1 in Fig.~\ref{fig:Kondologs}) had a significantly
smaller amplitude of  $\lsim 20 e^2 / h$, implying 
only about 10 active TLSs 
(that samples with a smaller
amplitude should show better scaling is to be expected
due to a smaller spread in $\Delta$'s, see
(Ti.3) in  section~\ref{sec:stressed}). 

\subsection[Open Questions in the 2CK Scenario]{Open 
Questions in the 2CK Scenario}
\label{openquestions}

Having discussed the successes of the 2CK scenario,
we now turn to questions
for which the 2CK scenario
is unable to offer a detailed explanation, namely
the conductance transitions (Cu.9), the strong
magnetic field dependence (Cu.8a), and
the microscopic nature of the TLSs. 
 We shall point out below
that (Cu.9) and (Cu.8a)  are   not generic to TLS-induced ZBAs,
and speculate that they are related and must involve some new
``high-energy'' physics, since (Cu.9) occurs at a large  voltage $V_c$.
Therefore, our lack of understanding of the latter need not
affect the 2CK interpretation
of the low-energy scaling behavior (Cu.6). 
We conclude with some speculations about the
microscopic nature of the TLSs, and the likelihood
that realistic TLSs will have all the properties 
required by the 2CK scenario.

\subsubsection{Conductance Transitions}
\label{sec:condtrans}

The fact that conductance transitions occur only in
samples that have a  ZBA (Cu.9a) suggests \cite{RB95} that these
are related to the ZBA:
if the latter is phenomenologically viewed as the
manifestation of some strongly correlated state of the system,
then   conductance transitions correspond to 
the sharp, sudden, ``switching off'' of the correlations
as $V$ becomes too large.
For example, in the 2CK interpretation,
interactions of electrons with TLSs in the constriction
give rise to  a strongly correlated  non-Fermi-liquid state
at small $T$ and $V$. One might speculate that
if for some reason a large voltage could ``freeze''
the TLSs, i.e.\ prevent  them from tunneling,
this would disrupt the correlations and
give rise to a sudden change in
the DC conductance and hence a spike in the
differential conductance. 

At present we are not aware of any detailed
microscopic explanation for the conductance transitions.
Note, though, that they do not occur in {\em all}\/
Cu samples showing ZBAs. Moreover, 
recent experiments by  Upadhyay {\em et al.}\/
\cite{ULB96}  on Titanium constrictions and  by 
 Keijsers {\em et al.}\/ \cite{KSvK95} on 
constrictions made from metallic glasses
showed TLS-induced ZBAs  with 
properties very similar to  RB's quenched Cu constrictions,
but no conductance transitions at all
(see (Ti.6) and (MG.4) in section~\ref{relatedexp}, where 
these experiments are reviewed).
This suggests that {\em conductance transitions
are   not a generic ingredient of the phenomenology
of ZBAs induced by TLSs.}\/
Moreover, in the quenched Cu samples,  provided that $H$
and $T$ are  sufficiently
small,  the transition voltage $V_c$
 at which the first conductance transition occurs usually lies
well above $\Tk$, the scale characterizing the extent of the 
low-energy scaling regime of the ZBA  [see Fig.~\ref{fig:dip+spike}(a)].
(In other words,  since they don't occur near zero bias,
the conductance transitions need not be
viewed as part of the zero-bias anomaly phenomenon at all,
if one restricts this term to refer only to the low-energy
regime.)

Thus, there seems to be a clear  separation of energy scales
governing the ZBA and the conductance transitions.
The latter must therefore be  governed by some new  large
energy scale due to a mechanism not yet understood.
However, due to the separation of energy scales, the
conductance transitions need not affect our description
of the low-energy scaling regime of the ZBA 
below $\Tk$ (which is $\ll e V_c/\kb$) in terms
of the 2CK model. 

\subsubsection{Strong Magnetic Field Dependence}
\label{strongmag}

Since the electron-TLS interaction is non-magnetic,
i.e.\ not directly affected by a magnetic field,
the 2CK scenario predicts no, or at best
a very weak magnetic field dependence for the ZBA.
This agrees with   the absence of a Zeeman splitting 
of the ZBA for the Cu samples (Cu.8b) (which 
was in fact one of the main reasons for the proposal
of the non-magnetic 2CK interpretation \cite{RB92}).
However, it leaves the strong magnetic field dependence (Cu.8a)
as a puzzle. 
(Two indirect mechanism for $H$ to couple to a 2CK system,
namely via $H$-tuning of the
asymmetry energy $\Delta_z (H)$ and via  channel symmetry breaking,
are investigated in Appendix~\ref{sec:magfield}; they
are found to be  too weak to account for (Cu.8a), 
contrary to the interpretation we had previously offered \cite{RLvDB94}.)

It is therefore very significant that 
the experiments by  Upadhyay {\em et al.}\/ on Ti constrictions
and by Keijsers {\em et al.}\/ on 
metallic-glass constrictions
 show  ZBAs with only a very weak or even no $H$-dependence
 [see section~\ref{relatedexp}, (Ti.5), (MG.3)], 
in complete accord with 2CK expectations. 
\label{sec:Handspikes}
This suggests that, just as the conductance transitions,
{\em the strong magnetic field dependence}\/
(Cu.8a) {\em of the quenched Cu constrictions is not a generic feature
of TLS-induced ZBAs.}\/ Moreover, Fig.~\ref{fig:condtrans}
suggests that in the Cu samples these
two properties might be {\em linked}\/, because
it shows that
 the strong $H$-dependence
of $G(V=0,H)$ is related to the fact that the
transition voltage $V_c$ decreases to 0 as $H$ is increased (Cu.9b,v).
(In other words, if the strongly correlated state sets in at 
smaller $V_c$ as $V$ is lowered, the voltage-regime $0< V<V_c$ 
in which the anomaly can develop 
is smaller, so that its total amplitude is smaller.)

Since the main difference between the
Cu constrictions and the Ti and metallic-glass constrictions seems to be
that the former contain TLSs with very small  $\Delta$'s
(see the next subsection),
whereas in the latter, being disordered materials, there will certainly
be a broad distribution of splittings,
we speculate that the conductance spikes
and strong $H$-dependence  might both be a consequence of
the very small $\Delta $s occuring in the Cu samples,
perhaps due to interactions
between several TLSs with very small splittings.

Thus, we conclude that attempts
(such as those in \cite{RLvDB94}) to explain
the $H$-dependence  of the ZBA (even at $V=0$)
purely in terms of the 2CK model, which
captures only the physics at low energies below $\Tk$,
are  misdirected, because  the $H$-dependence
would arise, via the conductance transitions,
from the  ``high-energy''
physics associated with the large scale $V_c$. 

This interpretation, according to which a magnetic field
does not directly affect the low-energy physics of the
phenomenon (only indirectly via its effect on $V_c$), 
can be checked by doing a $V/T$ scaling analysis
at fixed but small, non-zero magnetic field. If $H$ is sufficiently
small that the conductance transitions still
occur at relatively high voltages
(i.e.\ $V_c > \Tk)$, the scaling properties
of (Cu.6) should not be affected by having $H \neq 0$). 
The presently available data
is unfortunately insufficient to test this prediction.

The conclusions of this and the previous subsection
are summarized in assumptions (A3) and (A4) in 
subsection~\ref{summaryassumptions}.

\subsubsection{Microscopic Nature of the TLS}

Finally, the 2CK interpretation is of course unable to 
answer the question: What is the microscopic nature of
the presumed TLSs? Now, ignorance of
microscopic details does not affect our explanation
for why the scaling properties (Cu.6) of the ZBA
seem to be universal:
because the latter
are presumably governed by the {\em fixed point}\/ of the 2CK model,
any system that is somewhere in the vicinity
of this fixed point will flow towards it as the temperature is lowered 
(provided that relevant perturbations are
sufficiently small) and hence exhibit the same universal
behavior, irrespective of its detailed bare parameters.

However,  the quality of the scaling behavior 
implies some rather stringent restrictions
on the allowed properties of the presumed TLSs,
because we need to assume that all active TLSs
(e.g.\ about 10 for sample \#1, which shows the best scaling) 
 are close enough in parameter space to the \nfl\ fixed point 
to show pure scaling.

This implies, firstly, that 
interactions between TLSs (which are known to
exist in general  \cite{RB88}, mediated by
strain fields and changes in electron density),
must be negligible, because they would drive
the system away from the 2CK \nfl\ fixed point.
Secondly, the fact that scaling is only
expected in the regime $\Delta^2/ \Tk < T < \Tk$ can be used
 to estimate that  $\Tk \simeq 3$ to
5K and $\Delta \lsim 1$K (see section~\ref{ch:scaling} for 
details). 
Kondo temperatures  \label{sec:TK}
in the range of  1-10~K are in good agreement with the most
recent theoretical estimates for TLSs \cite{ZZ94}.
However, the condition  $\Delta \lsim 1$K
implies that for active TLSs the distribution
of energy splittings, $P(\Delta)$, must be peaked
below  $\Delta \lsim 1$K.
Since $\Delta
= (\Delta_z^2 + \Delta_x)^{1/2}$,
both the asymmetry energy $\Delta_z$ and 
tunneling rate $\Delta_x $ must 
be $ \stackrel{<}{\sim} $1K, 
a  value so small that it needs further  comment.

First note that  it is not immediately obvious
that values
of the bare tunneling rate $\Delta_x$ exist at 
all that allow 2CK physics:
For transitions to be able to take place, 
the barrier between the wells must be sufficiently
small,  but a small barrier is usually associated
with a large bare $\Delta_x$, implying a large bare $\Delta$
(and $\Delta$ sets the energy scale at which the
renormalization flow toward the \nfl\ fixed point is
cut off).
Now,  
for a TLS in a metal, the physics of screening 
can reduce the direct tunneling rate $\Delta_x$ 
by as much as three orders of magnitude under 
renormalization to $T \ll \Tk$ \cite{ZZ96a}
(when tunneling between
the wells, the tunneling center  has to 
drag along its screening clowd, which becomes increasingly
difficult, due to the orthogonality catastrophy, at lower
temperatures). Thus, the renormalized direct tunneling 
rate can always be assumed very small. Though this implies
a large effective barrier, it
does not necessarily prohibit 2CK physics:
 Zar\'and and Zawadowski \cite{ZZ94,ZZ94b} have  shown
2CK physics can be obtained even if $\Delta_x = 0$,
provided that the model contains some other channel for
inter-well transitions, such as 
electron-assisted  transitions
via more highly excited TLS states
(see Appendix~\ref{sec:excitedstates} of II).

More serious is the assumption that the renormalized
asymmetry energy $\Delta_r $
also be $ \lsim $1K.
This may seem very small when recalling that
in glassy materials,  the distribution $P(\Delta)$ for
the asymmetry energy is rather flat, with
$\Delta$  varying  over many (often tens of) Kelvins.
Note, though, that far from being glassy,
the constrictions are believed to be rather clean (Cu.5), containing
almost perfectly crystalline  Cu. Therefore,
our intuition about the properties of TLSs in glasses 
 can not be applied to the present system.
For example, the TLSs could possibly be  dislocation kinks.
(This would naturally account for the disappearance
of the ZBA when static disorder is added (Cu.4a),
since dislocation kinks can
be pinned by other defects.) 
Since the dislocation kink  would find itself 
in a rather crystalline material,
some lattice symmetry could guarantee then that the
two wells of the TLS are (nearly) degenerate and hence
assure a small $\Delta_z$ and hence small $\Delta$, etc.

 Moreover, some role  might be played by the mechanism of
``autoselection'': \label{sec:autoselection}
This assumes that a given TLS
will only be ``active'', in the sense of
contributing to the non-trivial $V$- and $T$-dependence
of the conductance, if its (renormalized) parameters
happen to be in the appropriate \nfl\ regime; if
they are not, the TLS would only be an ``inactive spectator''
that only affects the  $V$- and $T$-{\em independent}\/
background  conductance $G_o$. Moreover, provided
that the distribution $P(\Delta_z)$ is not zero
near $\Delta_z = 0$ (which seems very unlikely), 
there should always exist a few TLSs with $\Delta \lsim 1$K, 
since $\Delta_x$ is strongly reduced by screening.

%
 Despite the above arguments, the fact that
the effects of non-zero $\Delta$s did not show up in the
quenched Cu samples, necessitating 
the  assumption that all active TLSs must have $\Delta \lsim 1$K,   remains 
probably the weakest point the 2CK interpretation of these
samples. Therefore it is very significant that the
effects of non-zero $\Delta$s were recently explicitly 
demonstrated in the Ti constrictions of Upadhyay {\em et al.}\/:
in some samples scaling {\em breaks  down}\/ below an energy scale
associated with $\Delta$, which was found to be tunable by
electromigration and the application of a magnetic field
[see section~\ref{sec:stressed}, (Ti.4), (Ti.5)]. 
Thus, the new Ti experiments significantly bolster the 2CK scenario
at its hitherto weakest point.  

To shed further  light on the effects of non-zero
$\Delta$,  it would be interesting if 
ZBA experiments with {\em specific}\/ type of defects with 
{\em known}\/ parameters could be performed.

It should be mentioned that 
\label{sec:asymmetry}
Wingreen, Altshuler and Meir (WAM)
have recently claimed \cite{WAM95}(a) that 
the 2CK interpretation is internally inconsistent
if one takes into account the effect of static disorder:
using values for the coupling  constants deduced
from the observed Kondo temperatures, they concluded that
renormalized  energy splitting $\Delta $ would be dramatically
increased (to typical values of about 100K), 
and in particular that there would be 
zero probability for zero splitting ($P (0) = 0$). 
However, since their arguments  neglected the physics of screening
(i.e.\ the strong reduction of $\Delta_x$ under renormalization
to lower temperatures), we believe that their
conclusions, in particular that  $P (0) = 0$, are not persuasive
\cite{WAM95}(b), \cite{ZZ96a}.
A critical discussion of their arguments 
 is given in Appendix~\ref{sec:criticism} of paper II.

\subsection[Summary of Assumptions of 2CK Scenario]{Summary 
of Assumptions of 2CK Scenario}
\label{summaryassumptions}

The 2CK interpretation of the ZBA in quenched Cu constrictions
developed in this section can be summarized in
the following \vspace{2mm} assumptions: \\
\begin{itemize}
\item[(A1)]
 The ZBA is due to the presence in the constriction region
of structural defects, namely TLSs, that interact
with conduction electrons according to 
the non-magnetic Kondo model, which renormalizes at low energies
to the \nflr\ of the 2CK model.
\item[(A2)]
The TLSs may occur with a distribution of (dressed or
renormalized) parameters,
but all ``active'' TLSs, i.e.\ those
 which contribute measurably to the voltage
dependence of the conductance in the Cu point contacts,
must have parameters which cause their behaviors to be
governed by the physics of the \nfl\ fixed point of the
 2CK model.  This requires that all
of these active TLSs do not interact with each other.
Moreover, pure $V/T$ scaling will only occur in the window
$\Delta^2/ \Tk <  T < \Tk$ (in quenched Cu samples
it was found for $T > 0.4$K  and $\Tk > 3$ to 5K,
implying $\Delta \lsim 1$K). 
Note, though, that the absence of {\em scaling}\/ for
$T< \Delta^2/\Tk$ does not necessarily imply
the absence of non-Fermi-liquid physics,
which  can still show up as $V^{1/2}$ behavior
for $\Delta^2/\Tk < V < \Tk$.
\item[(A3)]
The large  magnetic field
dependence (Cu.8a) and the conductance transitions (Cu.9)
of the quenched Cu ZBAs are related, but
not generic to the ZBA.
They cannot be explained by 2CK physics alone, but
involve some new energy scale on the order of $V_c$. 
\item[(A4)] A magnetic field does not {\em directly}\/ influence
the low-energy physics ($V< \Vk$, $T< \Tk$) of the ZBA.
 Therefore, 2CK physics can account for the behavior of the ZBA
in quenched Cu samples at fixed  $H$, provided that
$H$ is sufficiently  small ($\lsim 1$~T)
that the conductance transitions do not
influence the 2CK scaling regime (i.e.\ $V_c > \Vk$).
\end{itemize}

\section{Scaling Analysis of $G(V,T)$}
\label{ch:scaling}

In this section, we present a detailed scaling analysis
of the data and establish the scaling properties of  (Cu.6). This is
a very important part of the analysis of the
experiment, since the scaling properties were used
above to eliminate quite a number of otherwise
plausible candidate explanations of the ZBA. Nevertheless,
since the upshot of this section is contained completely
in property (Cu.6), this section can by skipped
by readers not interested in the details of the scaling analysis.

Of course, the scaling properties (Cu.6) we are about to
establish are  simply experimental facts, independent
of any theoretical interpretation.
Nevertheless, during the writing of Ref.~\cite{RLvDB94},
these properties were predicted (before their 
experimental verification) on the basis of the CFT
 solution of the 2CK model, and we shall present our
analysis within this framework. We begin by giving
 in section~\ref{ch:scaling}.A the
general scaling argument first reported in \cite{RLvDB94} to motivate
the scaling Ansatz for the conductance $G(V,T)$,
and a back-of-the-envelope
calculation of the scaling function $\Gamma (x)$.
 A more careful calculation, tailored to the
nanoconstriction geometry and its non-equilibrium peculiarities,
is reserved for paper II.

\subsection[The Scaling Ansatz]{The Scaling Ansatz}
\label{sec:scalingansatz}

The 2CK model is known \cite{NB80}
to flow to a non-trivial, non-Fermi-liquid fixed
point at $T=0$, at which the model has been solved exactly
by Affleck and Ludwig (AL), using CFT
\cite{Lud94a}. This fixed point governs the physics in
the  \nflr, 
namely $\Delta^2/ \Tk <  T < \Tk$
and $V < \Vk$.   We shall now show by
a general scaling argument  \cite{RLvDB94} that  the
assumption of proximity to this fixed point (A2),
implies the scaling properties (Cu.6) of the conductance
$G(V,T)$. 

\subsubsection{General Scaling Argument}

Consider first the
conductance signal $G_i(V,T)$ due to a single TLS (labeled
by the index $i$) with $T \ll
T^i_{K}$, $eV \ll \kb T^i_{\sss K}$, $\Delta_i = 0$, but {\em arbitrary }\/
ratio
$eV/ \kb T$. According to the general theory of
critical phenomena, one expects that physical quantities will
obey scaling relations in the neighborhood of any fixed
point.
For the conductance in the present case,
a natural scaling Ansatz is:
\begin{equation}
\label{eq3.scalingansatz}
G_i (V,T) = G_{io} +  B_i T^{\alpha} \, \Gamma \!\left( {A_i eV
\over ( \kb T)^{\alpha/\beta}} \right) \, .
\end{equation}
The parameters $A_i$ and $B_i$ are non-universal, positive constants,
analogous  to the $a^{\pm}_i$ and $b_i$ of
\Eq{eq:2.diffconddef},  which may vary, for instance, as a
function of the position of the TLSs within
the constriction. However, the function
$\Gamma(v)$ should be a {\em universal function,}\/ a
fingerprint of the 2CK model
that is the same for {\em any }\/ microscopic realization
thereof. It
must have the asymptotic form $\Gamma(v) \propto v^{\beta}$ as
$v \to \infty $, since $G(V,T)$ must be independent of $T$ for $eV \gg
\kb T$.
Due to the arbitrariness of $A_i$ and $B_i$,
we are free to use the normalization conventions that
\be
\label{eq3.Gammanorm}
\Gamma(0) \equiv 1 \; , \qquad
\Gamma(v) = v^{\beta} + \mbox{constant} \quad
        \mbox{as}\; v^{\beta} \to \infty \; .
\ee

Now, {\em if $V$ is small enough}\/, its only effect will be to create a
slightly non-equilibrium electron distribution in the leads.
In particular, effects that directly affect the impurity itself,
like $V$-dependent strains, or the ``polarization''
 of the  TLS in one well due to
the non-equilibrium electron distribution, etc.\, can then be
neglected. In this case, which we shall call
the {\em weakly non-equilibrium regime,}\/
$V$ only enters in the Fermi functions
of the leads, in the form $[e^{\beta( \varepsilon - eV/2)}
+ 1]^{-1}$, i.e. in the combination $eV/\kb T$, implying \label{p:a=b}
$\alpha = \beta$.

For a constriction with several defects, the conductance
signal will be additive,\footnote{To be more precise: the
contributions of the impurities to the resistance $R=G^{-1}$
are additive, but since $R = R_o + \sum_i \delta R_i(V,T)$,
with $R_o  \gg \delta R_i(V,T)$, the form~(\ref{eq3.manydefects})
follows.}
 i.e. (now using $\alpha$=$\beta$):
\begin{equation}
\label{eq3.manydefects}
G(V,T)  = G_o  + T^{\alpha} \sum_i B_i \,
 \Gamma \!\left( {A_i eV \over  \kb T}
\right) \, .
\end{equation}
Subtracting $G(0,T)$ from this to  eliminate $G_o$ then
immediately results in the scaling relation of (Cu.6a):
\begin{equation}
\label{eq3.testscaling}
{G(V,T) - G(0,T) \over T^{\alpha}} \, = \, \sum_i B_i \left[ \Gamma
\!
\left( A_i v \right) - 1 \right] \, \equiv \, F(v)
\, .
\end{equation}
The scaling function $F(v)$ defined on the right-hand side 
is non-universal, since it depends on the $A_i$ and $B_i$.

This is as far
as general scaling arguments will take us;
a specific theory is needed to predict $\alpha$.
To this end,  we argue by analogy with the conductivity
of a {\em bulk}\/ metal containing 2CK impurities.
There the bulk conductivity $\sigma(T)$  is determined, via the Kubo formula,
\be
\label{eq3.Kubo}
        \sigma(T) =  2 {e^2 \over 3 m^2}
        \int \! {d^3 p \over (2 \pi)^3 }
        \left[ - \partial_{\varepsilon_p} f_o (\ve) \right]
         \vec p\, {}^2 \tau( \varepsilon_p) \; ,
\ee
by the elastic scattering life-time
$
\tau^{-1} (\ve) = - 2 \mbox{Im} \;  \Sigma^R (\ve) \; ,
$
where $\ve \equiv \varepsilon_p - \ve_{\sss F}$, and
$\Sigma^R (\ve)$ is the retarded electron self-energy.
$\Sigma^R (\ve)$ has been calculated exactly
by  Affleck and Ludwig, using CFT \cite{AL93},
for the bulk $k$-channel Kondo problem
in the neighborhood of its $T=0$ fixed point
(i.e.\ for $T \ll \Tk$). They
 found that in general $\tau^{-1}$ has the following scaling form
(motivated in  paper~III,
or \cite{AL93}, \cite[chapter~8]{JvDthesis}):
\bea
\label{eq3.tauscaling}
        \tau^{-1} ( \ve, \{
        \lambda_m \} ) &\equiv& - 2 \mbox{Im} \Sigma^R (\ve
        , \{ \lambda_m \}) \\
\nonumber
       &=& \tau^{-1}_o \left[ 1    + \sum_m \lambda_m \;  T^{\alpha_m} \;
        \tilde \Gamma_m (\ve / T) \right] \; .
\eea
The sum on $m$ is over all perturbations to the fixed point
action that one wants to consider. Each such perturbation
is characterized by a non-universal
parameter $\lambda_m$ which measures its  strength
(and has dimensions of $E_m^{-\alpha_m}$,
where $E_m$ is the  energy scale characterizing this perturbation),
a universal scaling dimension $\alpha_m$
and a (dimensionless) universal scaling function $\tilde \Gamma_m (x)$.
In principle all the $\alpha_m$ and
$\tilde \Gamma_m (x)$ (but not the non-universal $\lambda_m$)
can be calculated exactly from
CFT, provided all the $\lambda_m T^{\alpha_m}$ are small enough
that one is in the close vicinity of the fixed point.
Perturbations with $\alpha_m < 0 $ or $ > 0$ are
relevant or irrelevant, respectively,
because they grow or decrease as
the temperature is lowered at fixed $\lambda_m$.
For all perturbations of interest in this paper, the
scaling functions have the properties
$\tilde \Gamma_m (x) = \tilde \Gamma_m (-x)$ and
 $\tilde \Gamma_m (x) \propto x^{\alpha_m}$
as $x \to \infty$ (the latter property follows because
the perturbation must become $T$-independent in the limit $\ve >>
T$).

AL have calculated in detail
the {\em leading}\/ irrelevant correction to $\tau_o^{-1}$
for the $k$-channel Kondo problem,
for the case that no relevant perturbations are present.
In other words they take $\lambda_m = 0 $ for all $m$
for which $\alpha_m <  0$, and consider only
the correction corresponding to the smallest $\alpha_m > 0$,
say $\alpha_1$.
When referring only to this correction,
we shall drop the subscript $m=1$ and
denote the corresponding parameters by \label{lambda1}
$\lambda_1 \equiv \lambda$, $\alpha_1 \equiv \alpha$ and
$\tilde \Gamma_1 (x) \equiv \tilde \Gamma (x)$. They showed
that for the $k$-channel Kondo problem
$
        \alpha = {2 \over 2+k}$, $\lambda = \tilde \lambda \Tk^{- \alpha } > 0
$
(where $\tilde \lambda$ is a dimensionless
number of order unity) and $\tilde \Gamma (x) < 0$.\footnote{The sign of
$\lambda$ \label{f:tildeb} is {\em a priori}
undetermined in the CFT approach
(see \protect\cite[after eq.~(3.64)]{AL93});
however, to conform to the expectation that the Kondo scattering rate
increases as $\ve$ or $T$ are decreased, we need $\lambda > 0$,
since $\tilde \Gamma (x) < 0$.} (For an explicit expression
for $ \tilde \Gamma (x)$, see \cite[eq.(3.50)]{AL93}
or 
paper~II).

Considering only this leading irrelevant perturbation,
it follows immediately from the Kubo formula that
for the 2CK problem ($k=2$, hence $\alpha = \half$), the bulk
conductivity has the form
\be
\label{eq3.sigmascaling}
        \sigma (T) = \sigma_o + \left( { T \over T_{\sss K} }
        \right)^{1/2}   \sigma_1 \; ,
\ee
with $\sigma_1 > 0$. The unusual power law $ T^{1/2} $ is a signature of
the non-Fermi-liquid nature of the $T=0$ fixed point.
For a Fermi liquid, one would have had $T^2$.

Although the form (\ref{eq3.tauscaling}) for $\tau^{-1}(\ve)$
was derived by AL only for a bulk geometry,
it is natural to assume that it also governs
the conductance in the nanoconstriction geometry
(paper II is devoted to a careful justification 
of this assumption). This implies that the
exponent  in \Eq{eq3.scalingansatz} should also be $\alpha = \half$,
which completes our general-principles motivation for the scaling Ansatz.

\subsubsection{Back-of-the-envelope calculation of $\Gamma (v)$}
\label{sec:botenvelope}

If one is willing to gloss over important subtleties, it is
possible to obtain by a simple back-of-the-envelope calculation
a quantitative expression for the scaling function that
agrees with that found by more careful means in paper II.

Our starting point is \Eq{eq:2.diffconddef},
which gives the change in conductance due to back-scattering off defects
in a nanoconstriction in terms of the scattering rate $\tau^{-1}
(\ve)$. Now,
the main difference between a bulk metal and a nanoconstriction
is that the latter represents a decidedly non-equilibrium
situation. However, in the {\em weakly}\/
 non-equilibrium regime, i.e.\ if
 the voltage is small enough ($V < \Vk$), it is a reasonable guess
(which is substantiated in II) that the form of the  scattering rate
of electrons off a TLS in the nanoconstriction
is not all that different as when the TLSs are in the bulk.
Hence, let us boldly \label{p:boldly} use\footnote{
The justification for this assumption is explained in
\protect\sect{p:boldjust}  of paper II; essentially,
we assume that the leading non-equilibrium corrections to $\tau^{-1}$ are
of order $V/\Tk$, which are negligible  in the
weakly non-equilibrium or scaling regime.}
the {\em equilibrium}\/ form for $\tau^{-1}$, namely \Eq{eq3.tauscaling},
in \Eq{eq:2.diffconddef} for $\Delta G$, thus obtaining
(to lowest order in $\lambda_m$):
\bea
\nonumber
\lefteqn{        G(V,T, \{ \lambda_m \})  =  G_o
        - e^2/h  \sum_m \lambda_m T^{\alpha_m}  \!\!
         \int\!\! d \ve [- \partial_{\ve} f_o
        (\ve) ]} \qquad
\\ &&  \label{eq3.Gscaling}
\times \sum_i b_i
        \half  \Bigl[ \tilde \Gamma_m (\ve \!-\! \half e V a^+_i) +
        \tilde \Gamma (\ve \!+\!  \half e V a^-_i) \Bigr]
\eea
Now write $\ve/ \kb T  \equiv x$, $\; eV / \kb T \equiv v$,
$\; f_o (v) \equiv [e^v + 1]^{-1}$, and define  (universal)
functions $\Gamma_m(v)$ by:
\be
\label{eq3.defGamma}
         \gamma_{b, m} \Gamma_m (\gamma_{a, m} v) \equiv \int \! dx
        [ \partial_{x} f_o  (x) ]
        \tilde \Gamma_m (x + v/2) \; .
\ee
Here $\gamma_{a,m}$ and $\gamma_{b,m}$ are universal  constants,
chosen such that $\Gamma_m(v)$ is normalized as in \Eq{eq3.Gammanorm},
with $\Gamma_m (\infty) > 0$:
\be
\label{eq3.Gammannnorm}
        \Gamma_m(0) \equiv 1 \; , \quad
        \Gamma_m(v) \; = \; v^{\alpha_m} + \mbox{const., \quad as}
        \; \; v^{\alpha_m} \to \infty \; .
\ee
Using the property $\tilde \Gamma_m (x) = \tilde \Gamma_m (-x)$
in the first term of \Eq{eq3.Gscaling}, we find
\bea
\nonumber 
\lefteqn{        G(V,T , \{ \lambda_m \})  =  G_o
        +  e^2 / h \sum_m \lambda_m T^{\alpha_m} \gamma_{b, m}}
\\ && \label{eq3.Gprediction}
        \times  \sum_i b_i  \;
        \half  \Bigl[  \Gamma_m (a^{+}_i \gamma_{a, m} v) +
         \Gamma_m (a^{-}_i \gamma_{a,m} v) \Bigr] \; .
\eea

Let us now specialize again to the leading irrelevant
perturbation, for which only $\lambda_1 \equiv \lambda \neq 0$.
(Since for this case $\tilde \Gamma (x) < 0$,
 $\gamma_{a,1} \equiv \gamma_a$ and $\gamma_{b,1} \equiv
\gamma_b$ can both be chosen positive.)
In this case, \Eq{eq3.Gprediction}
 is precisely of the form
of the scaling Ansatz \Eq{eq3.manydefects}, with $\alpha = \half$.
Thus we have found
a ``derivation'' for the scaling Ansatz.\footnote{
Note though that \Eq{eq3.scalingansatz}
is actually a little too simplistic, since in
\Eq{eq3.Gprediction} each defect gives rise to two terms with
different $a^+_i$ and $a^-_i$.
Note also that $b_i$, $\gamma_o $ are by definition
all positive constants, and $\Gamma (x) > 0$.
With the choice $\lambda > 0$, discussed in footnote~\ref{f:tildeb},
we therefore have $G(V,T, \lambda) - G(0,T, \lambda) > 0$, consistent with
experiment.} Moreover, this little
calculation has furnished us with an expression,
namely \Eq{eq3.defGamma}, 
for the universal scaling function $\Gamma (v)$
in terms of the exactly
 known universal function $\tilde \Gamma (x)$.

This, in a nutshell, is all there is to the scaling
prediction. Of course, to back up this
result by a respectable calculation,
considerably more care is required and several
technical and conceptual hurdles have to be overcome. 
These are addressed in paper II. And finally, in paper III
the origin of the exponent $\alpha = \half$ is explained;
AL originally derived it using the full machinery of CFT,
but it in III it is found much more simply and directly using
 a recent reformulation \cite{ML95} of their theory in
terms of free boson fields.

\subsection[Scaling Analysis of Experimental Data]{Scaling 
Analysis of Experimental Data}
\label{sec:scaleexp}

In this section we summarize the results of a careful
 scaling  analysis of the
experimental data \cite{RLvDB94}, based on
section 6.4.2 of \cite{Ralph93}, and establish that
$G(V,T)$ has the properies summarized in (Cu.6) of
section~\ref{sec:data}.

\subsubsection{First Test of $T^{1/2}$ and $V^{1/2}$ Behavior}
\label{sec:sqrt}

As a first test of the scaling relation \Eq{eq3.manydefects},
one can consider it in the asymptotic limits $v \to 0$ and $\infty$ ,
in which the conductance becomes
[using \Eq{eq3.Gammanorm}]:
\bea
\label{eq3.limitT}
        G(0,T) &=& G_o + T^{\alpha}  B_{\Sigma} \; ,  \\
\label{eq3.limitV}
        G(V,T_o) &=& const +    v^{\alpha}   F_o
        \quad  \mbox{at fixed }\: T_o \ll eV/ \kb \; ,
\eea
where $ B_{\Sigma}  \equiv \sum_i B_i$ and 
$  F_o \equiv \sum_i B_i A_i^\alpha$. 
 Figs.~\ref{fig:6.7}(a) and~\ref{fig:6.8}(b) confirm that
$G(0,T)$ and $G(V, 0)$ roughly conform to \Eqs{eq3.limitT}
and~(\ref{eq3.limitV}), respectively, with $\alpha = \frac12$.
Values for $B_{\Sigma}$ and $F_o$ can be obtained from
straight-line fits to these data,
and are listed in table~\ref{tab:constants}.
However, the quality of these data
is not good enough to rule out other values
of $\alpha$, ranging from 0.25 to 0.75.

\subsubsection{Scaling Collapse}
\label{sec:collapse}

A much more stringent determination of $\alpha$
can be obtained from  the scaling properties
of the  combined $V$ and $T$ dependence of the $G(V,T)$,
which, according to the 2CK interpretation, should
follow \Eq{eq3.testscaling}.
To check whether the data obey this relation,
the left-hand side of  \Eq{eq3.testscaling}
should be plotted vs.\ $v^{\alpha}$.
Provided that  the correct value of $\alpha$ has been chosen,
 the low-$T$  curves for a given
sample should all collapse,  with no further adjustment of free
parameters,  onto the sample-specific
scaling curve $ F(v)$ vs. $v^{\alpha}$, which should be linear
for large $v^{\alpha}$ [by \Eq{eq3.limitV}]. By adjusting $\alpha$ to
obtain the best possible collapse,  $\alpha$ can be determined
from the data rather accurately. The 2CK scenario
 predicts $\alpha = \half$.

The raw data for the
differential conductance $G(V,T)$ of sample \#1 of Fig.~\ref{fig:6.8}
is\label{p:heating}  shown
in \fig{fig:6.10}(a), for $T$ ranging from 100~mK to 5.7~K.
Using $\alpha = \half$, rescaling as in \Eq{eq3.testscaling} and plotting
the left-hand side vs. $v^{1/2}$, these data have the form
shown in \fig{fig:6.11}(b). The data at low $V$ and low $T$ collapse
remarkably well onto one curve,
which we shall call the
{\em scaling curve}.\/ Furthermore, $F(v)$ vs. $v^{1/2}$
has linear asymptote as $v \to \infty$, in agreement with
\Eq{eq3.limitV}.

Most of the individual curves deviate from the scaling curve
for $V$ larger than a typical scale $e \Vk \simeq 1$meV
(this is also roughly the voltage at which lower-lying
curves in Fig.~\ref{fig:6.10}(a) begin to
fall on top of each other) . Likewise,
the lowest curves in the figure,
which  correspond to the highest temperatures,
deviate from the scaling curve for almost all $V$.
These deviations from scaling at high $V$ and $T$ are
expected, since if either $V> \Vk $ or $T > \Tk $,
the scaling Ansatz is expected to break down.
We estimate $\Tk$ as that $T$ for which the rescaled
data already deviate from the scaling curve at
$eV/\kb T \le 1$. This gives $\Tk \ge 5$~K for the defects of
sample~1, \label{p:TK} which are reasonable values, as discussed in
section~\ref{sec:TK}, and establishes the empirical relation
$e \Vk \simeq 2 \kb \Tk$.

A somewhat complimentary estimate for the highest Kondo
temperatures of the TLSs in these samples comes from the temperature
at which the zero-bias signals first become visible as the samples
are cooled. For all 3 samples featured here, this value is
approximately 10~K.

The quality of the scaling provides an exacting test of the
exponent in the scaling Ansatz. Using $\alpha = 0.4$ or $0.6$
instead of 0.5 in  \Eq{eq3.testscaling} produces a clear
worsening of the collapse of the data (see Fig.~2(b) of \cite
{RLvDB94}).
%
As a quantitative measure of the quality of scaling,
we define  the parameter $D (\alpha)$, which is the mean square deviation
from the average scaling curve $\bar F (v) \equiv \sfrac{1}{N} \sum_{n=1}^N
F_n (v)$ (where $n$ labels the different experimental
curves, corresponding to different temperatures $T_n$),
 integrated over small values of $|v|$:
\be
\label{eq3.dev}
        D (\alpha)
         \equiv \sfrac{1}{N} \sum_{n=1}^N \int_{- v_{max}}^{v_{max}}
        dv \left[ F_n (v) -  \bar F(v) \right]^2 \; .
\ee
$D(\alpha) = 0$ would signify perfect scaling.
Taking the 5 lowest $T$ ($\leq 1.4$~K) and $v_{max} = 8$
(these are the data which {\em a priori}\/ would be expected to
be most accurately within the scaling regime, since
they are closest to the $T=0$ fixed
point), one obtains \fig{fig:deviations}(a). Evidently the
best scaling of the data requires $\alpha = 0.48 \pm 0.05$
[the estimated uncertainty of $\pm 0.05$ comes
from the uncertainty in the exact minimum in
the curve in \fig{fig:deviations}(a)]. This is in remarkably
good agreement with the CFT prediction of $\alpha = \half$.

 We have also tested the more general scaling
form of Eq.~(\ref{eq3.scalingansatz}), and have observed
scaling for $0.2 < \alpha < 0.8$, with
$(\beta- 0.5) \approx (\alpha - 0.5)/2 $, with best scaling for
$\alpha = 0.5 \pm 0.05$. But as argued earlier on
page~\pageref{p:a=b}, one expects $\alpha
= \beta$ on general grounds.

The scaling Ansatz has also been tested on two other Cu samples.
The rescaled data for sample~2 [Fig.~\ref{fig:6.14}(a)] collapse well
onto a single curve at low $V$ and $T$, for
$\alpha = 0.52 \pm 0.05$ [\fig{fig:deviations}(b)]
and with $\Tk \ge 3.5$~K.
At high $V$ and high $T$ the non-universal  conductance spikes of (Cu.9)
are visible.  The data for sample~\#3 do not seem to collapse as
well [Fig.~\ref{fig:6.15}(b)] (illustrating how
impressively accurate by comparison the scaling is for
samples~\#1 and~\#2).
However, we suggest that this sample in fact displays two separate
sets of scaling curves (see arrows),
one for $T \le 0.4$~K and one
for $0.6\;\mbox{K} \le T \le 5$~K, with interpolating curves in between.
This could be due to defects with a distribution of $\Tk$'s, some
having $\Tk\simeq 0.4$~K and others having $\Tk \ge 5$~K. The second
(higher-$T$) set of curves do not collapse onto each other as well
as the first, presumably because there is still some (approximately
logarithmic) contribution from the $\Tk \simeq 0.4$~K defects.

\subsubsection{Universality}
\label{sec:universality}

If for any sample all the $A_i$ in Eq.~(\ref{eq3.testscaling}) were
equal, one could directly extract the
universal scaling curve from the data. The curve obtained by plotting
\be
\label{howtoplot}
        { G(V,T) - G(0,T) \over B_{\Sigma} T^{1/2}} \quad
        \mbox{versus} \quad
        (A eV/ \kb T)^{1/2}\; ,
\ee
with $A$ determined by the requirement that the
asymptotic slope be equal to 1 [compare \Eq{eq3.Gammanorm}],
would be identical to the
universal curve $(\Gamma(x)-1)$ vs.\  $x^{1/2}$. Such plots are
shown in
Fig.~\ref{fig:6.16}(b).  The fact that the scaling curves for all three
samples are indistinguishable indicates that the distribution of $A_i$'s
in each sample is quite narrow and is a measure of the
universality of the observed behavior.

To make possible quantitative comparisons of the data with
the CFT prediction of \Eq{eq3.defGamma},
we now proceed to extract from the data the value of
a universal (sample-independent)
constant
[essentially a Taylor
coefficient of $\Gamma(v)$], which is
independent of the possible distribution
of $A_i$'s and $B_i$'s.


Consider the regime $v\gg 1$. As argued earlier, here $\Gamma(v)
\simeq v^{\beta}$, and since $\beta$=$\alpha$=1/2, with
the normalization conventions of \Eq{eq3.Gammanorm} we can write,
asymptotically
\begin{equation}
\label{etafunction}
\Gamma(v) -1 \equiv v^{1/2} + \Gamma_1+ O(v^{-1/2}) \, ,
\end{equation}
where $\Gamma_1 $ is a universal number. It characterizes how
long the $\Gamma(v)$ curve stays ``flat'' for 
small $v$ before bending upwards towards its asymtotic $v^{1/2}$-behavior.
It follows from
Eq.~(\ref{eq3.testscaling})
that
\begin{equation}
\label{Gsigmaapprox}
F (v) =  v^{1/2} F_0 + F_1 + O(v^{-1/2}) \, ,
\end{equation}
where $F_0 \equiv \sum_i B_i A_i^{1/2}$ 
and $F_1 \equiv \Gamma_1 B_{\Sigma}$.    
Values for $F_0$ and $F_1$ may be determined
from the conductance data by plotting $F$
versus $v^{1/2}$ and fitting the data for large
$v^{1/2}$  to a straight line.  For samples~1
and~2 we fit between $(eV/\kb T)^{1/2}= $ 2 and~3, and for sample~3
(using only the curves below 250 mK) between
2 and 2.5.

Values for $F_0$, and
$F_1$ are listed in Table~\ref{tab:constants}. The uncertainties
listed are standard deviations of values determined at different
$T$ within the scaling regime for each sample.
From these  quantities, an experimental determination of
the universal number $\Gamma_1= F_1/B_{\Sigma}$ can be obtained;
it is listed in Table~\ref{tab:constants}. The values of
$\Gamma_1$ are consistent among all~3 samples,
in agreement with our expectation that $\Gamma_1$ should be
a universal number.

Paper II will be devoted to a calculation of the universal
scaling curve $\Gamma (v)$ and the universal number $\Gamma_1$.
There it is shown that the quantitative predictions
of the 2CK scenario are indeed consistent with the measured
scaling curve.

\subsection[Upper Bound on the Energy Splitting $\Delta$]{Upper 
Bound on the Energy Splitting $\Delta$}
\label{sec:Delta}

The energy splitting $\Delta$ of a TLS
is a relevant
perturbation to the degenerate 2CK fixed point.
In the language of
the magnetic Kondo problem, it
acts like a local magnetic field, and hence has
scaling dimension $-\half$
(see (3.19) of \cite{AL92b},
or \cite[section 3.4.1 (e)]{CZ95}).
Therefore a  non-zero $\Delta$ implies that
the electron scattering rate $\tau^{-1} (\ve)$
and hence the conductance $G(V,T)$
will contain  correction terms (to be labelled by $m\!=\! 2$),
given by Eqs.~(\ref{eq3.tauscaling}) and
(\ref{eq3.Gprediction}), respectively, with
$\alpha_2 = - \half$ and $\lambda_2 = {\tilde \lambda_2
\Delta/ E_2^{1 \over 2}}$.  Here $\tilde \lambda_2$
is a dimensionless number of order unity, and $E_2$ is an energy that
sets the scale at which $\Delta$ becomes important.
Though this scale  is not {\em a priori} known in AL's CFT treatment,
we shall take $E_2 = \Tk$, since no other obvious energy
scale suggests itself.

It follows that such  $\Delta$-dependent corrections,
which would spoil good $V/T$ scaling,
are unimportant only if $\lambda_2 T^{\alpha_2} < 1$, i.e.\
as long as
\be
\label{DeltaTTk}
        \Delta <   (T \Tk)^{1 \over 2}
\ee
holds for each active TLS. Note that  this inequality
allows $\Delta$ to be somewhat
larger than the naive estimate that would follow
from $\Delta <  T$. As emphasized by Zawadowski \cite{Zawpriv},
this somewhat enlarges the window of parameter space
in which the 2CK  scenario is applicable.

The above analysis enables us to estimate
an upper bound on the energy splittings of active TLSs occuring
in the quenched Cu samples. The data for samples~1 and~2
show  pure $T^{1/2}$ behavior at $V=0$
(i.e.\ absence of $\Delta$-corrections)  for $T$ as small as 0.4~K.
Taking $\Tk \!\approx\!5$~K, \Eq{DeltaTTk}
implies that for any active TLS,
$\Delta  \!<\! 1.4$~K. This upper bound is rather small,
and was discussed at some length in section \ref{sec:autoselection}.

\section{Related Experiments}
\label{relatedexps}
\label{relatedexp}

In recent years, ZBAs have been found in a number of
different nanoconstriction studies
\cite{ULB96,KSvK95,KSvK96,AG93}.
It should be appreciated that in each the  ZBA could in principle 
be caused by a different mechanism. However, two recent experiments
have found ZBAs that convincingly seem to
be of the same type and origin as those in the quenched Cu
constrictions. We review their properties below in the form
of a continuation of the list of properties
 compiled in section~\ref{sec:data}, together with their
interpretation in terms   of the 2CK scenario.

\subsection[Titanium Nanoconstrictions]{Titanium Nanoconstrictions}
\label{sec:stressed}

Three of us   (ULB) \cite{ULB96}  
have recently studied nanconstrictions
with the same geometry as the quenched Cu constrictions of RB, but with the 
leads made from Titanium (Ti). This is a
stressed refractory metal, which is both more disordered and in
a state of higher tensile stress than Cu, and thus is a 
likely candidate to have dynamical defects. The following properties 
were found:

\begin{itemize}
\item[(Ti.1)] {\em General Properties:}\/ 
\\
(a) ZBAs occur in more than 90\% of the samples. ---
This high rate of occurence
is due to the highly stressed nature of refractory metals
(mean free path is estimated to be $l \gsim 10$~nm),\footnote{
Note that the mean free path is not much larger (if
at all) than the typical constriction radius (5- 15 nm). Strictly 
speaking, this means that
one is approaching the regime in which the theory of diffusive,
not ballistic, point contact spectroscopy should be used
(section~\protect\ref{sec:pcspectroscopy}), but  we shall continue
to use the latter.} 
and the consequent abundance of TLSs.\\
(b) The typical amplitude is about $\Delta G \simeq 10 e^2/h$. ---
This indicates that just a few (probably less than 5) 
TLSs are involved, since the 2CK model implies a maximum
$\Delta G$ of $ 2 e^2/h$ per defect (see footnote~\protect\ref{2e2h}). 
\\
(c) The ZBA anneals away at room temperatures on a time
scale of a few days to a few months. --- This is significantly longer than
in Cu samples, because Ti has a higher melting temperature.
\\
(d) {\em Geometry-induced stress:}\/ 
If a dirty insulating substrate is used (e.g. with organic contaminants),
to which the Ti-film does not stick well, the ZBAs were absent in
almost all the samples. The ZBA occurs only if there is good
adhesion between metal film and substrate. 
Nevertheless, the ZBA is {\em not  a surface effect}\/
(e.g.\ due to TLSs on the surface caused
by a mismatch in lattice constants between Ti and the substrate). 
This  was demonstrated as follows by studying constrictions made from
a combination of  Palladium (Pd) and Ti: 
Samples made purely from Pd
almost never had ZBAs; even when they  did,
the ZBA annealed away  over a period of a few hours, 
in contrast to the much longer annealing times for Ti (Ti.1c). 
This implies that stresses are relieved much quicker in Pd than in Ti. 
Now, for  a series of  samples, first
a layer of Ti of thickness 2, 5, 10
or 25nm was deposited on the both sides of the constriction,
and thereafter Pd was used to fill up the bowl and form the leads. 
If the ZBA were a surface effect, it should have shown up in these
samples.  However, very few of them had ZBAs,  and even when they did,
the ZBA annealed away  over a period of a few hours,
exactly as for the pure-Pd samples. 
However when the Ti layer's thickness was increased
above  30nm, which in these samples 
is approximately the cross-over thickness for filling
the bowl sufficiently to  form a continuous nanobridge
through the constriction,  ZBAs started to appear
with annealing times characteristic of Ti. This 
shows that the ZBA's cause is situated in the bulk of the bowl 
and not on its surface. --- This revealing investigation implies 
that {\em the ZBA results from geometry-induced stress in the metal}\/.
This stress  is provided by the bowl-like shape of the hole
(but only when good adhesion to the bowl is possible)
and anneals away very slowly in Ti, but rapidly in Pd.
This conclusion strongly supports the assumption (A1) of
section ~\ref{summaryassumptions} that the ZBA 
is caused by stress-induced structural defects.

\item[(Ti.2)] {\em $V/T$ scaling:}\/ 
Some samples show  the same scaling behavior,
$[G(V,T) - G(0,T)]/T^\alpha = F(eV/\kb T)$, as that observed
for Cu samples (Cu.9a). Of those samples for which
a scaling analysis according to  section~\ref{ch:scaling}
was done, at 
least three scale very well for $\alpha \simeq 0.5$;
scaling plots for one of these (called sample~4 here) 
are shown in Fig.~\ref{shashi1}.
\\
--- This is in excellent agreement with the 2CK scenario, 
given assumption (A2)
that the TLS energy splittings $\Delta$ are sufficiently small.
The scaling curve has the same universal shape as for the Cu constrictions
(compare with Fig.~\ref{fig:6.10}), with  the universal number
$\Gamma_1 = -0.81 \pm 0.10$ [see \Eq{etafunction} and 
Table~\ref{tab:constants}],  but actually scales
better, in that the ``bending-down'' deviations from scaling
 occuring in Fig.~\ref{fig:6.11}(b) 
for Cu at large $(V/T)^{1/2}$ are {\em absent}\/ here.
This makes it somewhat hard to unambigously determine
the Kondo temperature; estimates give $T_K \simeq 10 - 20$K. 
\item[(Ti.3)] {\em Premature $V^{1/2}$ saturation:}
Other samples, such as sample~5 of Fig.~\ref{shashi2},
 show for small $T$ a ``premature'' saturation of the $V^{1/2}$-behavior,
implying a very marked breakdown of  scaling. This 
can be quantified as follows:
The conductance (also for  sample~4) can be fitted, at fixed $T$, by the 
phenomenological form 
\be
\label{Gstressed}
        G(V) = G(0) + a [V^2 + (2 T_x)^2]^{1/4} \; ,
\ee
where $T_x$ characterizes the ``saturation energy'' at which 
 the large-voltage $V^{1/2}$-behavior crosses over to the flat low-$V$ regime.
For the scaling sample~4 of (Ti.2) 
one finds $T_x \simeq T$
[this is the reason for including the phenomenological
factor of 2 before $T_x^2$ in  \Eq{Gstressed}],
meaning sample~4's  saturation is due to thermal rounding.  
In sharp contrast, for sample~5  one finds that 
for sufficiently low temperatures, 
 $T_x$ ($=\! 1.43 \! \pm \! 0.03$K)  is
{\em much larger}\/ than $T$ (by a factor of almost 20(!) for the 
lowest $T$ of 76mK). This implies ``premature  (non-thermal)
saturation'' of $V^{1/2}$ behavior  as $V$  is lowered. 
(No such saturation was ever observed in Cu samples
down to 50 mK).\\
---
The 2CK scenario attributes [see assumption (A2)] 
premature  $V^{1/2}$ saturation
to the presence of some TLS with
finite energy splittings $\Delta  \simeq T_x$. At energies
below $\Delta$, 2CK physics ``feezes out'', and
{\em pure $V/T$ scaling
  behavior}\/ is destroyed for  $(T T_k)^{1/2} < \Delta$
[\Eq{DeltaTTk}], due to the presence of a new energy scale.
Note, though, that  non-Fermi-liquid physics is destroyed
only at low energies --- it becomes observable again 
 at sufficiently high energies, as demonstrated by the reemergence of 
$V^{1/2}$ behavior {\em above}\/ $T_x$, in agreement with 
the last sentence of assumption (A2).
Importantly, this implies that 
even if a constriction contains a large number
of impurities with a wide distribution of $\Delta$'s,
it {\em can}\/  show $V^{1/2}$ behavior at sufficiently large
$V$, though its scaling will suffer. 
 (Indeed, for the Cu samples the quality of scaling
was better (Cu.2) for samples with  smaller ZBA amplitudes,
i.e.\ fewer TLSs.)
\item[(Ti.4)] {\em Electromigration:}\/ 
\\
(a) Electromigration, the application of large voltages ($V$=200 mV,
$J\simeq 10^{10} \mbox{A/cm}^2$) 
for short periods of time (10 seconds, several times), can cause 
significant changes in the saturation energy $T_x$. 
For sample~5, Fig.~\ref{shashi3}(a) shows that  $T_x$ changed
from 2.3 to 1.4K as a result of such an electromigration. 
\\
(b) Another  device (sample~6)
had a more complicated low-$V$ behavior [Fig.~\ref{shashi3}(b)], 
characterized by
a sum of two terms of the form (\ref{Gstressed}), with 
two distinct saturation energies $T_{x1}$ and $T_{x2}$. 
Upon electromigration, they experienced (opposite!) changes,
from 0.9 to 1.5K and 9.7 to 6.8K, respectively. 
\\
--- Presumably electromigration, which is known to controllably cause
defect rearrangement \cite{Rallsthesis,RRB89},
 modifies the parameters of some TLSs, thereby changing their $\Delta$s. 
(Ti.4b)  is direct evidence that {\em individual}\/ defects
are responsible for the ZBA; 
evidently, two different TLSs dominate the ZBA in this particular case.
\item[(Ti.5)] {\em Magnetic field:}\/ 
\\
The magnetic field dependence of $G(V,H)$ is weak and random
(see Fig.~\ref{shashi4}): when $H$ is changed from 0 to 5 T, $G(V,H)$
\\
(a) changes by less than $2 e^2/h$, with random sign, 
for  $V <  T_x$;  and\\
(b) is completely $H$-independent for  $V >  T_x$.
\\
--- The very weak $H$-dependence is consistent with 2CK expectations
(see Appendix~\ref{sec:magfield}). It shows
that the strong $H$-dependence observed for Cu ZBAs (Cu.8a) is not
a generic property of the ZBA,  as discussed in section~\ref{strongmag}
and summarized in assumptions (A3,A4). 
For Ti constrictions, the entire $H$-dependence can be attributed to 
{\em $H$-tuning of $\Delta(H)$,}\/ i.e. to the
fact that in disordered materials, 
the TLS splitting is known \cite{ZGH91,GZC92} to be a 
random function $\Delta (H)$ of $H$ due to disorder-enhanced
interference effects (see section~\ref{slowfluctuators}). 
$H$-tuning of $\Delta(H)$ is consistent with
the fact that the $H$-dependent changes in $G$ have roughly
the same size (Ti.5a) as those due 
to changes in $\Delta$ induced by electromigration (Ti.4).
It  explains the random sign of the magnetoconductance (T1.5a),
and also  explains (Ti.5b), because $\Delta (H)$ can only affect
$G(V)$ {\em below}\/ $T_x$. 
\item[(Ti.6)] {\em No conductance transitions:}\/
The conductance transitions (Cu.9) that occured in at least 80\%
of the quenched Cu samples have never been observed in
any of the Ti samples. 
--- This shows that Cu conductance transitions (Cu.9) 
are not  generic, as discussed  in section~\ref{sec:condtrans}
and summarized in assumption (A3).
\end{itemize}

The above discussion shows that the  Ti nanoconstrictions 
display all the phenomenology expected from 2CK impurities:
(i)  their amplitudes are sufficiently
small to be attributed to only a very few TLSs; 
(ii) some of them show good $V/T$ scaling with scaling exponent
$\alpha = \half$;
(iii) others demonstrably show the
effects of finite, tunable $\Delta $; and (iv)  they lack the
puzzling conductance transitions and large
magnetic field dependence of the quenched Cu samples. 
Thus, they appear to be ``custommade''
realizations  of 2CK physics in nanconstrictions.

\subsection[Mechanical Break Junctions Made from Metallic 
Glasses]{Mechanical Break Junctions made from Metallic Glasses} 
\label{mbj}

Keijsers, Shklyarevskii and van Kempen \cite{KSvK96}
studied ZBAs in  mechanically controlled break junctions
 made from metallic glasses, which are certain to contain many TLSs.
They observed the following properties:

\begin{itemize}
\item[(MG.1)] {\em Amplitude and shape:}\/ \\
The ZBA has  qualitatively exactly
the same shape and sign as that of RB's quenched Cu samples,
with an amplitude of sometimes more than 100~$e^2/h$. ---
The large amplitude is to be expected, since metallic glasses
contain a high  concentration of TLSs.
\item[(MG.2)] {\em Slow ZBA fluctuations:}\/ \\
(a) Remarkably, in some samples
the amplitude of their ZBA fluctuates  between
two values $G(V)$ and $G' (V)$ (or sometimes several)
in a telegraph-noise fashion on a time scale
of seconds, evidently due to the presence of one
(or sometimes several) slow fluctuators
(see section~\ref{slowfluctuators}) in the constriction
region. 
\\
(b) The amplitude of these telegraph-fluctuations,
$\Delta G (V) = |G - G'|$, is of order $1e^2/h$ or less,
{\em and depends on}\/ $V$. It decreases from
$\Delta G(0)$  to 0 as
$V$ increases from 0 to between 5 and 10 mV, see Fig.~\ref{fig:zarand}.
\item[(MG.3)] {\em Magnetic field:}\/ \\
The ZBA shows no  $H$-dependence. ---
This is exactly as expected in the 2CK scenario
(see section~\ref{strongmag}), and confirms the conclusion
(Ti.5) that the strong $H$-dependence of Cu ZBAs
 (Cu.8a) is not a generic property of the ZBA. 
\item[(MG.4)] {\em No conductance transitions:}\/
The conductance transitions (Cu.9) of the quenched Cu samples 
have never been observed  metallic-glass constrictions
\cite{privateK}. 
--- This confirms the conclusion (Ti.6) that Cu 
conductance transitions (Cu.9)  are not  generic.
\end{itemize}

The $V$-dependence of $\Delta G(V)$
implies that the large features of the ZBA and
the small amplitude fluctuations cannot be  unrelated phenomena.
For example, it is not possible 
 to attribute the overall
ZBA to a suppression in the density of
states due to static disorder (analogous to the proposal
of WAM for quenched Cu samples, 
see Appendix~\ref{p:WAMWAM}), while assuming
the additional small conductance fluctuations  to be caused by
an independent slow fluctuator. The problem with 
such a scenario would be that the amplitude of the fluctuations,
though of the right magnitude of $< e^2/h$, would be
$V$-{\em in}\/dependent.

Keijsers {\em et al.} state that the large features
of their ZBA can be explained  by invoking either
Zawadowski's non-magnetic Kondo model (section~\ref{non-mag})
or KK's theory of TLS-population spectroscopy (Appendix~\ref{KKpop})
to describe the interaction of electrons with the {\em fast}\/
TLSs in their system. They propose that the  amplitude fluctuations 
can be explained (in either theory) by
assuming that the TLS-electron interaction strengths
[$V^z$ and $V^x$ in \Eq{HTLS}] of  some fast TLSs
are {\em modulated between two values}\/,  due to short-ranged interactions 
with a nearby slow two-state system, when the latter
 hops between its two positions. 

Zar\'and, von Delft and Zawadowski recently pointed out
\cite{vDZZ97} that the 
maximum  switching amplitudes  observed by KSK are so small 
 ($\Delta G_{max} < 1 e^2/h$ for all samples in Ref.~\cite{KSvK96})
that they seem to stem from  the parameter-modulations,
induced by a slow fluctuator, 
of only one or two fast TLS, where the parameter expected
to be most strongly modulated is the TLS asymmetry energy,
which can be assumed to fluctuate between two values,
$\Delta_z \Leftrightarrow \Delta_z'$. This implies
that the experiments of KSK constitute the first measurements of 
the conductance contributions of {\em individual}\/ fast TLSs,
and allows an unprecedentedly detailed comparison with theory:
by calculating  $ G (V, \Delta_z)$ (the contribution to the 
ZBA due to scattering off a TLS with asymmetry energy $\Delta_{z}$)
 for various asymmetry
energies, it should be possible to
find two values $\Delta_{z}$ and $\Delta'_{z}$ for
which $| G (V, \Delta_z)-  G (V, \Delta_z')|$ 
reproduces the measured $\Delta G(V)$.

Analysing two samples in detail,
Zar\'and, von Delft and Zawadowski showed that $\Delta G (V)$
could not be fit using the TLS-population spectroscopy
theory of Kozub and Kulik (section~\ref{KKpop}). However, 
rather good fits (Fig.~\ref{fig:zarand}) were achieved
using Zawadowski's non-magnetic Kondo model
(although the analysis does not completely rule out
that there can be a small KK contribution \cite{vDZZ97}).

This constitutes
possibly the most direct evidence yet for the applicability
of the 2CK  model to TLS-induced ZBA's in point contacts. 
Subsequent measurements of the response of such break junctions 
to RF-irradiation \cite{KSvK97} support this
conclusion (although they, too, do not completely rule
out a  small KK contribution). If both the  $V$- and $T$-dependence
of $\Delta G$ were known, a $V/T$ scaling analysis
\cite{RLvDB94} would serve as a further test of
the non-magnetic Kondo scenario. (At present the
 break junctions are not sufficiently stable against
mechanical  deformations under  temperature changes
to reliably determine the $T$-dependence of $\Delta G$.)

\section{Conclusions}
\label{conclusions}

\subsection[Summary]{Summary}
\label{sec:scalingsummary}

This paper is the first in a series of three (I, II, III) devoted
to 2-channel Kondo physics.
We have reviewed in detail the
experimental facts pertaining to a possible realization
of the 2CK model, namely the non-magnetic ZBA
in quenched Cu nanoconstrictions, and also integrated
into our analysis insights obtained from new experiments on
Ti and metallic-glass constrictions. 

  We have summarized the various experimental facts 
 for the quenched Cu samples in the form of nine properties, (Cu.1) to (Cu.9)
 (section~\ref{sec:data}).
Properties (Cu.1-5), which are of a mainly qualitative nature
and very robust, place very strong demands on any
candidate  explanations of the ZBA:
the zero-bias anomalies disappear under annealing,
and hence must be due to {\em structural}\/ disorder;
they disappear when static disorder is intentionally added,
and hence cannot be due to static disorder -- instead
they must be due to {\em dynamical}\/ impurities;
they show no Zeeman splitting in a magnetic
field (Cu.8b), and hence must be of {\em non-magnetic}\/ origin.
These observations lead to
the proposal \cite{RB92} that  the zero-bias anomalies
are due to nearly degenerate two-level systems, interacting
with conduction electrons according to the
non-magnetic 2-channel Kondo model of
Zawadowski \cite{ZZ94},  which renormalizes at low energies
to the \nflr\ of the 2CK model.

We then presented a quantitative analysis
of the $V/T$ scaling behavior of the conductance $G(V,T)
= G_o +  T^\alpha F(eV/\kb T)$, which demonstrates unambigously
that the scaling exponent has the unusual value of $\alpha = \half$,
in contrast to the usual Fermi-liquid value of $\alpha = 2$.
We argued that this too can naturally
be understood within the phenomenology of the $T=0$ fixed point of
the 2CK model, within which the experimental verification
of $\alpha = \half$ constitutes the direct observation of
a non-Fermi-liquid property of the system.
Breakdown of scaling for larger $T$ and $V$ values
is explained too, since for these
the system is no longer fine-tuned
to be close to the $T=0$ fixed point, thus spoiling
the scaling behavior. Estimates of $T_{\sss K}$
in the range 1-5~K, which is reasonable, were obtained,
as well as  an upper bound for the energy splitting of all active
TLSs of  $\Delta \lsim \kb 1$K.
This bound is rather small
(and has been criticised, see section~\ref{sec:asymmetry}),
but is enforced by the quality of the observed scaling.

The scaling analysis provides sufficiently detailed
information about the low-energy physics of the
system that it enabled us to rule out several other 
candidate mechanisms for explaining ZBAs.
(An alternative interpretation of the scaling
properties recently proposed by Wingreen, Altshuler
and Meir can be discounted on other grounds,
see  Appendix~\ref{p:disorder}). 

We then reviewed experiments on  Ti and metallic glass samples 
in the form of further  properties (Ti.1) to (Ti.6) and 
(MG.1) to (MG.4). They provided further strong support
for the 2CK interpretation: the Ti data demonstrated the
destruction of scaling in the presence of a non-zero, tunable
TLS energy splitting $\Delta$, and the metallic glass data
allowed the contribution of a {\em single}\/ TLS to be measured
and compared with theory.




The 2CK interpretation
is sufficiently  successful in accounting for
the observed phenomenology of the scaling properties (Cu.6),
that we believe more quantitative calculations based on
this model to be  justified.
The remaining two papers in this series,
II and III, are devoted to a quantitative calculation
of the scaling function $\Gamma (v)$, to be
compared with the experimental curve in \fig{fig:6.16}(b).
The final result is shown in section \ref{ch:calc} of
paper II, \fig{fig:scalingcurve}.
When our results are combined with recent numerical results
of Hettler {\em et al.} \cite{HKH94,HKH95}, quantitative
agreement with the experimental scaling curve is obtained.

The main conclusion of this investigation is therefore that
{\em the 2CK interpretation is in qualitative and quantitative
agreement with the scaling properties of the data.}\/
 It can also account for all other observed properties, 
with only two exceptions for the quenched Cu samples, summarized below.

\subsection[Open Questions and Outlook]{Open Questions and Outlook}
\label{furtherexp}

There are two experimental observations for the quenched Cu samples
that do appear to lie beyond the present
understanding of 2CK physics:
the conductance transitions (Cu.9) and
the apparantly related  strong magnetic field dependence (Cu.8a).
A theory of this phenomenon would be most welcome.
However, these effects appear to involve either ``high energy'' effects
or effects due to interactions between nearly degenerate TLSs which
are beyond the scope of the present-day single-impurity calculations 
of the 2CK model, which are applicable only to the 
low $T$- and $V$-regime  in the 
neighborhood of a $T=0$ fixed point.
However, these two experimental effects should not be considered 
generic to the physics of TLSs in nanoconstrictions:
they are absent in metallic-glass and Ti constrictions,
and particularly the latter, 
which  display both scaling and
the destruction thereof by a non-zero, tunable $\Delta$,
seem to be almost ``ideal'' realizations of 2CK physics.

\label{theoryquestions}

On the theoretical side, the 2CK model has recently
been subjected to renewed scrutiny (catalyzed in part by
its application to the ZBA and the claim that non-Fermi-liquid
behavior has been observed). The main point of contention is
whether any  realistic TLS would ever flow towards the
\nfl\ fixed point of the 2CK model, because
of the inevitable presence of relevant perturbations
that drive the system away from this point. 

Wingreen, Altshuler and Meir \cite{WAM95} have argued that
static disorder could lead to a significant
asymmetry energy $\Delta$ between the two states of the TLS
(a relevant perturbation). We
 critically discuss their arguments \cite{WAM95}(b), \cite{ZZ96a} in
Appendix~\ref{sec:criticism} of paper II, and judge them
not to be persuasive. More recently,
studying a  formulation of
the model that is slightly different from that
introduced by Zawadowski,
Moustakas and Fisher \cite{MF96} have discovered another relevant
operator (which was then interpreted by
Zawadowski {\em et al.} \cite{Zawprep} to 
be due to particle-hole symmetry
breaking). However, their conclusions have themselves been
questioned in Ref.~\cite{Zawprep,Ye96}, where the prefactor
of this new relevant operator was estimated to be negligably small,
and  for other technical reasons, 
some of which are mentioned in Appendix~\ref{sec:MF} of paper
II.


Certainly, further theoretical work is needed to fully understand
the stability, or lack thereof, of the $T=0$ fixed point of the
degenerate 2CK model.  Both experimental and theoretical work
would be welcome to better understand the nature of the defects
giving rise to ZBAs in metal point contacts, and the parameters
governing these defects.  Skepticism of the 2CK interpretation of
the data is not unwarranted, since this is seemingly an exotic
effect.  However, this model has provided a rather complete
account of the experimental observations (Cu.1-7), along with
accurate predictions of the scaling properties of the conductance
signals as a combined function of $T$ and $V$.  No other existing
model, based on more familiar physics, has been able to account
for all the data.

{\em Acknowledgements:}
It is a pleasure to thank B. Altshuler, D. Cox,
 D. Fisher, S. Hershield,  M. Hettler, R. Keijsers,
Y. Kondev,  V. Kozub,  J. Kroha, 
A. Moustakas,  A. Schiller, G. Sch\"on, H. van Kempen, 
G. Weiss, N. Wingreen, I. Yanson,
G. Zar\'and  A. Zawadowski for discussions. This work was 
partially supported  by the National Science Foundation
through Award No. DMR-9407245, the MRL Program, Award No. DMR-9121654, and 
the Cornell MSC DMR 9632275;
and by the Office of Naval Research, Award No. ONR N00014-97-I-0142.
A.W.W.L. and D.C.R. 
acknowledge financial support from the A.P. Sloan Foundation.
The research was performed in part at the Cornell Nanofabrication
Facility, funded by the NSF (Grant. No. ECS-9319005), Cornell University,
and industrial affiliates.

\appendix

\section{Ruling out Some Alternative Interpretations}
\label{sec:elimination}

In this appendix we discuss 
a number of conceivable  explanations  for the ZBA that could
come to mind as possible alternatives to the 2CK scenario.
We argue that each is inconsistent with some of
the experimental facts (Cu.1) to (Cu.9) and hence can be ruled out. 
Most of this material is contained in \cite{Ralph93}, \cite{RB95} and
\cite{WAM95}. Nevertheless, since some of these
arguments have been the subject of some controversy \cite{WAM95},
they deserve to be restated and summarized here,
for the sake of completeness and convenience.

\subsection{Static Disorder}
\label{p:WAMWAM}

Could the ZBA be due to static disorder?\label{p:disorder}
For example, one could consider attributing the decreased conductance
near  $V=0$ to either  weak localization due to disorder \cite{Berg84} or
disorder-enhanced electron-electron interactions
\cite{LR85}. In fact, the latter possibility
(first mentioned, but deemed implausible,
in Ref. \cite{HKH94}), was recently advocated \cite{WAM95} by
Wingreen, Altshuler and Meir (WAM)
(these authors  also offer a critique \cite{WAM95} of a crucial
assumption of the 2CK scenario, which we discussed in
section~\ref{sec:asymmetry}).

WAM made the interesting observation that if just the region of
the device near to the point-contact orifice were highly disordered,
this would give rise to a local depression
in the density of states near the Fermi surface
of the form
$
        \delta N(\ve - \ve_{\sss F},T)
        \propto - T^{1/2}
        F \left( {\ve - \ve_{\sss F} \over  T} \right)
$,
where $F$ is a scaling function.
This in turn would reduce the rate at which electrons incident
ballistically into the disordered region could traverse the
sample.
The total conductance would hence be reduced by an amount
$
        \Delta G(V,T) \propto
        \partial_V \int_{\ve_{\sss F}}^{\ve_{\sss F} + eV} \! d \ve
        \, \delta N(\ve - \ve_{\sss F}, T)
        = \delta N(V,T) .
$
Due to the scaling form of  $\delta N$,
this argument explains the scaling property (Cu.6),
and in fact the
scaling curve $F(v)$ of \Eq{eq:2.scalingform} that it produces
is in quantitative agreement with that of sample~1
(see \cite[Fig.1]{WAM95}). According to their estimates,
this scenario would require a disordered region of diameter
50~nm (the size of the bowl), a mean free path $l= 3$~nm
and a diffusion constant $D= 15$~cm${}^2$/s, i.e. rather
strong disorder.

The WAM scenario is appealing in that it accounts for the unusual
$T^{1/2}$ behavior using well-tested physical ideas,
without having to evoke any exotic new physics
(such as 2CK non-Fermi liquid physics).  However, it is at odds with
a number of qualitative (and hence very robust)
properties of the ZBA \cite[section 6.6.1]{Ralph93},\cite[(b)]{WAM95}:

\begin{enumerate}
\item According to (Cu.4a), upon the intentional introduction
of  static disorder the
ZBAs are not enhanced, as one would have expected in WAM's
 static disorder scenario, but {\em disappear
completely,}\/ in contradiction to the latter.
\item
The quenched Cu constrictions actually are considerably {\em cleaner}\/
than   is assumed in WAM's scenario,
as can be seen from three separate arguments:
\\
(a) According to (Cu.5), a direct estimate
of the mean free path, based on the point contact phonon
spectrum (a reliable and well-tested diagnostic method
\cite{JvGW80,DJW89}) suggests $l \gsim 30$~nm instead of WAM's 3~nm.
\\
(b)
WAM attempted to explain (Cu.3a),
 the disappearence of the zero-bias anomalies
under annealing, by assuming that the presumed static disorder
anneals away at room temperature.
However, this suggestion fails a simple quantitative
consistency check: let us model the constriction region
by a Cu cylinder 40 nm in diamter and 40 nm long, with
$l = 3$~nm. Estimating the resistance of this cylinder
using the Drude model yields $R=7 \Omega$, which
would be the dominant part of the resistance of the
device ($R < 10 \Omega$ in the lower-$R$ devices).
If annealing now removes sufficient disorder
that  the ZBA disappears, $l$ would have to increase considerably,
implying that the overall $R$ of the device would necessarily
decrease by tens of percent, which  contradicts (Cu.3b)
(according to which resistance sometimes even increases
 under annealing).
\\
(c) According to (Cu.4b) even a somewhat
smaller amount of disorder ($l = 7$~nm) than assumed by WAM
has been observed to cause
voltage dependent conductance fluctuations due to quantum
interference. However, these tell-tale signs of static
disorder were never seen in the quenched
ZBA samples (Cu.4c), though they did appear as soon as
disorder was  purposefully induced using electromigration (Cu.4b).
In other words, in Cu nanoconstrictions the signature of
static disorder is conductance fluctuations, not a ZBA.
\item
In the static disorder scenario,
the conductance depends
only on the {\em average}\/ disorder in the bowl
(not on the precise configuration of individual defects). Therefore,
it is unclear how to account for  the complex behavior
of the ZBA under thermal cycling (Cu.3c), under 
electromigration (Ti.4), and for the $V$-dependence of
the slow ZBA fluctuations $\Delta G(V)$ observed
in metallic glass break junctions (MG.2). 
Particularly the latter two facts clearly demonstrate
 that the ZBA strongly depends on {\em individual} defects. 
\item
The static disorder scenario predicts a $H^{1/2}$ behavior for
the magnetoconductance, and hence is at odds with 
the very weak $H$ dependence (Ti.5) of the Ti ZBAs. 
\item
The static disorder scenario provides no hint
at all about the possible origin of the
conductance transitions. WAM have suggested
that these may be due to superconducting
regions in the constriction (caused by
an attractive electron-electron interaction
at short range), but this suggestion
fails to account for the presence of
several different transitions in the same sample
(moreover, superconducitivity in a Cu sample
seems highly implausible).

\end{enumerate}

\subsection{Magnetic Impurities}

The asymptotic dependence of the conductance on
$\ln V$ or $\ln T$ (Cu.7) 
is reminiscent of the magnetic Kondo effect, where the
resistance increases as $\ln T$ with decreasing $T$
(as long as $T > T_K$).
However, there are at least three strong arguments that rule out magnetic
impurities as the source of the anomalies:
\begin{enumerate}
\item An effect due to magnetic impurities would not
anneal away at higher temperatures (Cu.3a),
since magnetic impurities are stable within constrictions,
not annealing away at room temperatures over a time scale of
6 months \cite{RB95}.
\item If the magnetic Kondo effect were at work,
a magnetic field would cause a well-known Zeeman splitting
in the zero-bias conductance dip, as
has been observed in nanoconstrictions intentionally
doped with the  magnetic impurity Mn \cite[section 5.2]{Ralph93},
as shown in  Fig.~\ref{fig:magnetic}(a).
However, in the devices under present consideration,
a Zeeman splitting has never been observed (Cu.8b).
\item Magnetic impurities in metal break junctions have been observed
to cause ZBAs that do not exhibit splitting because the Kondo
temperature scale is larger than the Zeeman energy \cite{Yanson95}.
However the
ZBA signals caused by these impurities are very different than the ones
we investigate, because they exhibit Fermi liquid scaling ($\alpha=2$)
rather than the $\alpha=0.5$ we measure.
\end{enumerate}

\subsection{TLS Population Spectroscopy}
\label{KKpop}

Many of the qualitative features of the quenched Cu ZBAs 
can be understood
within the framework of Kozub and Kulik's (KK)
theory of TLS-population spectroscopy \cite{KK86,KSvK95},
which has recently been extended by Kozub, Rudin, and
Schober \cite{KRS95}.
This theory assumes that the constriction contains
TLSs  (labelled by $i$)
with non-zero energy splittings $\Delta_i$,
so that the application of a voltage will induce a non-equilibrium
population $n_{i \pm} (V)$ of the higher and lower
 states $|\pm \rangle_i$ of
each TLS. Assuming that these two states 
have different 
cross-sections $\sigma_i^\pm$ for scattering electrons,
the resistance $R (V)$ will then depend non-linearly 
on the voltage and temperture. According to Kozub and Kulik,
the differential resistance has the form 
\be
\label{KKdRdV}
        {1 \over R} {dR \over dV}
        = \sum_j {e C_j \over 2 \Delta_j} (\sigma_j^+ - \sigma_j^-)
        \tanh ( \half \tau^{-1})
        S(\nu_j, \tau_j, q_j) \; ,
\ee
where $\nu_j = e V/\Delta_j$, $\tau_j = \kb T/\Delta_j$
and $C_j$ and $q_j$ are
geometrical coefficients depending on the  location of the 
$i$-th TLS in the constriction.
The function $S(\nu, \tau, q)$, which
they calculated explicitly, determines the shape of the
differential resistance curve (see Fig.~2 of \cite{KK86}),
which can vary quite significantly, depending on
the parameters $q$ and $\tau$. Note also that since
the signs of $ (\sigma_j^+ - \sigma_j^-)$ are arbitrary
(except under special assumptions, see \cite{KR96}),
\Eq{KKdRdV} predicts that ZBAs of both signs should occur.

The shape of the  ZBAs  measured by RB is qualitatively
the same as that predicted by KK's theory when the latter is averaged 
over many impurities 
(KK's theory has significant freedom for curve-fitting, due
to the undetermined parameters $q_j$ and $\Delta_j$).
However, since the function depends on the two parameters
$\nu$ and $\tau$ separately, this theory cannot
account for the existance of a scaling law
found in (Cu.6a), and certainly not for the specific value
$\alpha = \half$ of the scaling exponent (Cu.6b). 
Moreover, as pointed out in 
section~\ref{relatedexp}, 
recent related experiments by Keijsers {\em 
et al.} \cite{KSvK96}, in which the behavior of 
{\em individual}\/ fast TLSs were probed, shows that
they can not be reconciled with KK's theory \cite{vDZZ97} (see also
 \cite{KSvK97}).


\subsection{Properties of External Circuit}

It was pointed out to us by G. Sch\"on \cite{SSS95} that
fluctuations in the voltage $V$ due to
fluctuations in the external circuit can be shown
to lead to a conductance $G(V,T)$ that satisfies
the $V/T$-scaling behavior in (Cu.6) (but
with the exponent $\alpha$ determined
by the external resistance of the circuit,
and hence non-universal).

However,
since the ZBAs only occur in quenched samples (Cu.1), and
since ZBAs anneal away at room temperature (Cu.3a),
they must be due to some internal properties
of the sample. Hence they cannot be due to properties of
the external electrical circuit, such as
external voltage fluctations.

\subsection{Charge Traps and Other Possibilities}

The insulating material used in the devices, namely
amorphous Si${}_3$Ni${}_4$, may contain charge traps \cite{Mil71},
which could act as Anderson impurities or quantum dots
through which conduction electrons could hop.
This could cause dips in the differential conductance
through several mechanisms, such as Kondo scattering
from Anderson impurities \cite{MWL93}, inelastic hopping conduction
\cite{GM88,XMB90} or Coulomb blockade effects \cite{AL91}.

However, charge traps can be ruled out for the
present experiments for the following reason.
A charge trap has in fact been unambiguously observed in
a different experiment by Ralph and Buhrman \cite{RB94}.
The conductance shows a very characteristic
{\em peak}\/ at $V=0$, in complete contrast to the ZBA-dip.
The suggestion of Ref.~\cite{RB94} that this is a Kondo peak
that can be associated with
Anderson hopping of electrons through the trap
was taken up by K\"onig {\em et al.}
\cite{Koenig}, who calculated the conductance
$G(V,T)$ for this scenario and found reasonably good
agreement with that experiment.
 In other words,
if charge traps are present, their signals are unmistakable,
and very different from the ZBAs of present interest.

Other reasons ruling out charge traps as causes for the 
ZBAs may be found in Ref.~\onlinecite{Ralph93}, Section 6.6.2. 
Also in  Ref.~\onlinecite{Ralph93}, Section 6.6, a number of other
mechanisms were also considered and ruled out as causes
 for the observed ZBAs:  electronic surface states or quasi-localized states
within the metal, defect rearrangement, mechanical instabilities,
superconducting phases and heating effects.

\section{Magnetic Field Dependence in 2CK Scenario}
\label{sec:magfield}

It was stated in section~\ref{strongmag}
that (contrary to the interpretation we had previously offered
in Ref.~\cite{RLvDB94})
2CK physics is unable to account for
the strong magnetic field dependence (Cu.8a) of 
the ZBA. To illustrate this, we now 
investigate the two most obvious  mechanims
through which the 2CK scenario could conceivably produce an
$H$-dependence for the ZBA. These are the $H$-tuning of the
asymmetry energy $\Delta_z (H)$, and  channel symmetry breaking.
Both  drive the system away from the
degenerate 2CK fixed point (but not in precisely the same manner),
so that $H$ enters as a relevant perturbation.
However, we shall conclude that both mechanism are too weak
to explain the strength of the observed $H$-dependence in
quenched Cu samples.

\subsection{$H$-Tuning of $\Delta$}

One  possible mechanism by which
$H$ could couple to the system is 
by tuning \cite{ZGH91,GZC92} the TLS asymmetry energy  $\Delta_z (H)$,
and hence the energy splitting $\Delta(H)$, which
are then random functions of $H$
(see section~\ref{slowfluctuators}). 
With $\Delta (H)$ as a relevant perturbation, 
the analysis of section~\ref{sec:Delta} applies directly,
and a correction to the conductance proportional to
$\Delta (H)$ can be expected.

However, in this scenario, the magnetoconductance $G(H)$ should be a
random function of $H$ (since $\Delta(H)$ is), whereas it seems
to be always positive for the samples investigated in more detail.
Note also that it  would be incorrect to attribute the  non-universal
non-monotonic features seen at large $H$ for sample~\#2
to the random behavior of $\Delta(H)$, since
closer scrutiny reveals that this behavior is due to the $H$-motion
of the conductance transitions (Cu.9b,v).
Moreover, since $H$-tuning of $\Delta$ has
its origin in quantum interference, it is expected
to occur mainly in strongly disordered environments,
which the Cu samples are decidedely not [see (Cu.5)].
Furthermore, it would cause conductance changes 
of order $2 e^2/h$ per TLS
substantially  smaller than those observed (Cu.8a)
(particularly since the signs of the conductance changes
for different TLSs are random, leading to partial cancellations).

Hence, it seems as though $H$-tuning of $\Delta$
is not consistent with the observed $H$-dependence of
the quenched Cu samples.

\subsection{Channel Symmetry Breaking by $H$}
\label{sec:magphen}

The second mechanism by which a magnetic field could affect
a 2CK system is Pauli paramagnetism, which
{\em breaks channel symmetry}\/ (recall that the
channel index $\sigma$ refers to the Pauli spin $\uparrow, \downarrow$)
by causing a net magnetic moment $M = \mu_{\sss B}^2 H
N(\varepsilon_{\sss F})$ \cite[eq, (10.11)]{Zim72}.
Any such symmetry-breaking term can in principle give
corrections to the critical behavior, and should hence
be included in the CFT analysis.

A channel-symmetry breaking field is known to be a relevant
perturbation with scaling dimension $-\half$,
(see eqs.~(3.15) of \cite{AL92b}).
Hence, in direct analogy to our analysis of
the effects of $\Delta$ in section~\ref{sec:Delta},
it causes a perturbation (to be denoted by $m=3$) to the conductance
 $G(V,T,H)$,  described by \Eq{eq3.Gprediction} with
$\alpha_3 = - \half$ and $\lambda_3 = {\tilde \lambda_3
\mu_{\sss B} |H| / E_3^{1 \over 2}}$. Here $\tilde \lambda_3$
is a dimensionless number of order unity, and $E_3$ is an energy that
sets the scale at which $|H|$ becomes important.
Only the absolute value of $H$ enters, because the model is
otherwise symmetric in spin $\uparrow$, $\downarrow$,
so that the sign of the channel-symmetry-breaking field
cannot be important.

This correction term in \Eq{eq3.Gprediction}
implies that at a fixed, small temperature $T_o$
and $V=0$, the conductance obeys
\be
\label{eq:Hchannel}
        G(0,T_o, H) - G(0,T_o, 0) \propto |H| \; .
\ee
As was argued in Ref.~\cite{RLvDB94},
the available data is at least qualitatively not
in contradiction with this prediction,
since Fig.~\ref{fig:6.17} shows
non-analyticity at $H = 0$  and an  initial roughly
linear behavior (Cu.8a) (note though, that $H^{1/2}$ behavior
can not be ruled out either).

What are the effects of channel-symmetry breaking at sufficiently
large $H$? Presumably, the polarization of
the Fermi sea will become so strong that one channel
of conduction electrons  (the one with higher Zeeman energy)
will decouple from the impurity altogether,
and the system will cross over\footnote{The cross-over behavior
between the fixed points [e.g. the behavior of $G(0,T,H)$]
can not be calculated from CFT, which can only describe
the neighborhood of fixed points;  it might be possible, though,
to calculate this function exactly using Bethe-Ansatz
techniques.} to the one-channel
Kondo fixed point, at which the conductance $T$-exponent
is $\alpha=2$. Hence, at this fixed point the conductance,
 at fixed, large $H$, should obey the $V/T$
scaling relation \Eq{eq3.testscaling}, with $\alpha = 2$
\cite[eq.~(D29)]{AL93}.

However, it seems unlikely that these considerations 
of the large-$H$ regime have any
relevance at all for the Cu samples, since at large
magnetic fields, the conductance transitions have
moved into the ZBA-regime (Cu.9b,v), presumably destroying
all remnants of universal
2CK physics. (Indeed, a scaling analysis
for sample~\#2 at fixed $H=6$~T shows best scaling at neither $\alpha =
\half$ nor 2, but at $\alpha = 0.3$ \cite{RB95}, though this value
probably does not have special significance either.)

Having investigated the phenomenology to be expected
from channel symmetry breaking, let  us
 now step back and estimate the likely magnitude of this effect.
The Pauli paramagnetism that causes channel symmetry
breaking  merely shifts
the {\em bottom}\/ of the spin-up Fermi sea
relative to that of the spin-down Fermi sea by 
$\mu_{\sss B} H$, while their respective Fermi-surfaces remain aligned.
Therefore, the magnitude of the effect that $H$
has on the Kondo physics near the Fermi surface will
be of order $\mu_{\sss B} H / D$ (where
$D \sim \ve_{\sss F}$ is the bandwidth),
and hence negligible. Though this
argument is not conclusive
(e.g.\  in poor-man scaling approaches the band-width is
renormalized to much smaller values of order
$D'=\mbox{max}[V,T,\Delta]$),
it casts serious doubts on
the  channel-symmetry breaking scenario, in
particular because the observed amplitude of the magnetoconductance
is by no means small.

Thus, we have to conclude that 2CK physics cannot account for
the observed $H$-behavior (Cu.8a). In
section~\ref{strongmag} it was therefore suggested to be
linked to the $H$-motion of the conductance transitions
[$V_c(H) \to 0 $ as $H$ increases, (Cu.9b,v)].

\newpage 

\newpage 

\newpage 

\newpage \begin{figure}
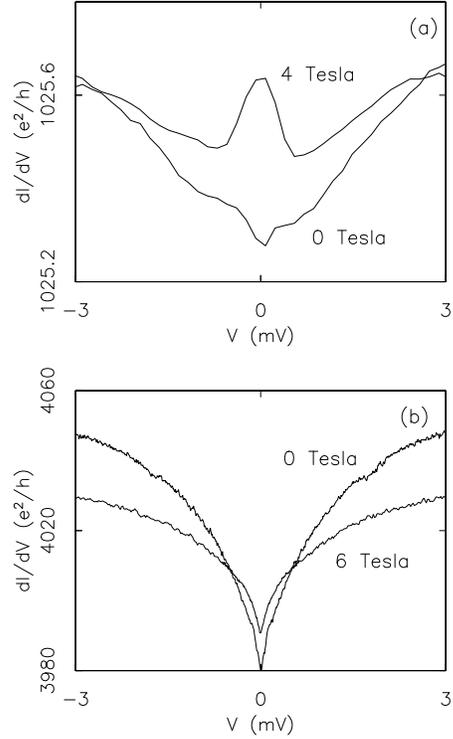
 
\vphantom{.}
\vspace{4cm}
\centerline{\psfig{figure=janfig4a.ps,height=6cm}}
\vphantom{.}
\vspace{3.7cm}
\centerline{\psfig{figure=janfig4b.ps,height=6cm}}
\vphantom{.}
\vspace{-2cm}
\caption[Zeeman splitting of conductance signals
for magnetic impurities in applied magnetic field.]{
(a) Conductance signals for 500 ppm magnetic Mn impurities
in Cu at 100~mK, showing Zeeman splitting in an applied magnetic
field. 
(b) The ZBA signals from quenched Cu samples
exhibit no Zeeman splitting, demonstrating that they are not due to
a magnetic impurity. However, the shape and amplitude of the
ZBA does depend on magnetic field. 
\label{fig:magnetic}}
\end{figure}

\newpage \begin{figure} 
\vphantom{.}
\centerline{\psfig{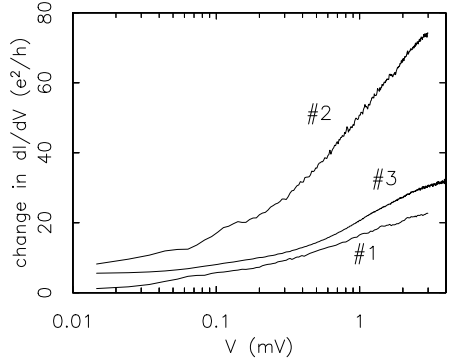}}
\vphantom{.}
\vspace{-01.1cm}
\centerline{\psfig{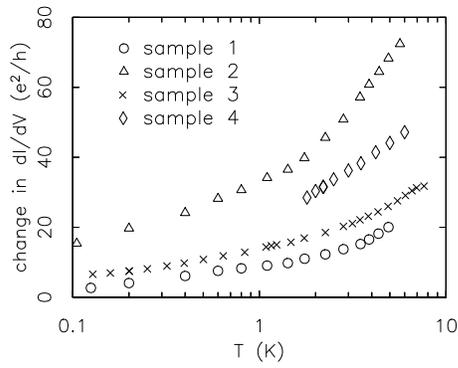}}
\vphantom{.}
\vspace{1cm}
\caption{
(a) $V$-dependence of the differential conductance
for $B=0$ and $T= 100$~mK, for three different 
samples \#1, \#2, \#3.
(b) $T$-dependence of the conductance for
$B=0$ and $V=0$ for the three samples of (a), and a fourth. 
\label{fig:Kondologs} }
\end{figure}

\newpage \begin{figure}
\vphantom{.}
\vspace{1cm}
\centerline{\psfig{figure=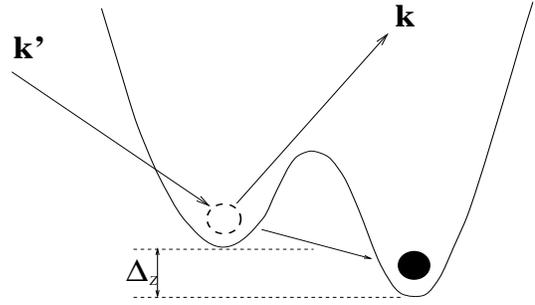,height=4cm}}
\vphantom{.}
\vspace{2cm}
\caption{
\label{fig:doublewell} A generic two-level-system,
with (bare) energy asymmetry $\Delta_z$ and tunneling rate $\Delta_x$.
An electron-assisted tunneling event is depicted:
an electron scatters of the TLS and induces the atom to tunnel.}
\end{figure}


\newpage \begin{figure}
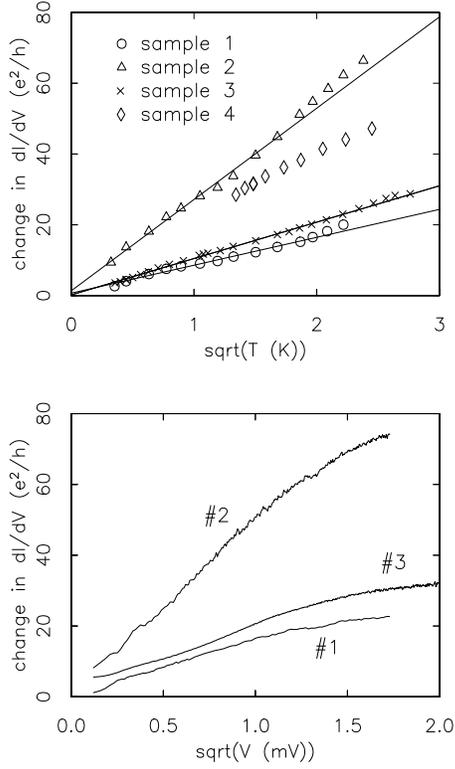

\centerline{\psfig{figure=janfig7a2.ps,height=6cm}}
\vphantom{.}
\vspace{-1.1cm}
\centerline{\psfig{figure=janfig7b2.ps,height=6cm}}
\vphantom{.}
\vspace{1.1cm}
\caption{
(a) Temperature
dependence of the $V=0$ conductance $[G(0,T)- G(0, T_o)]$ for the 4
unannealed Cu samples of Fig.~\protect\ref{fig:Kondologs}, plotted
versus $T^{1/2}$. The values of $G(0,T)$ for the different
samples, extrapolated to $T= 0$ as shown are for
sample~\#1: 2829 $e^2/h$, sample~\#2: 3973 $e^2/h$,
sample~\#3: 30.8 $e^2/h$, and sample~\#4: approximately 2810 $e^2/h$.
\label{fig:6.7}
(b) Voltage dependence of the differential conductance
at $T=100$~mK for some of the same samples as in (a),
plotted versus $V^{1/2}$.
The size of deviations from $T^{1/2}$ behavior in
(a)  (1 part in 3000) is consistent with the
magnitude of amplifier drift in these measurements, as they were
performed over several days. The $V$-dependent measurements in (b)
are less subjective to such drift problems, as they are
taken over a much shorter time span.
\label{fig:6.8}}
\end{figure}

\newpage \begin{figure}
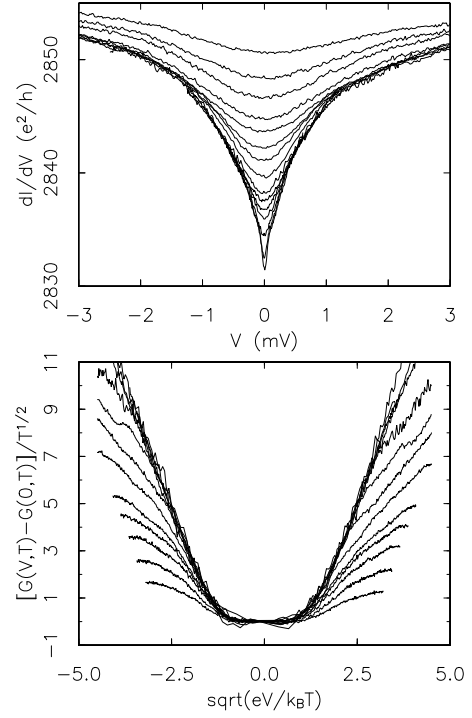

\vphantom{.}
\vspace{1cm}
\centerline{\psfig{figure=janfig8a.ps,height=6cm}}
\centerline{\psfig{figure=janfig8b2.ps,height=6cm}}
\caption{
(a) Voltage dependence of the differential conductance
for sample~\#1 of Fig.~\protect\ref{fig:6.7}\protect, plotted
for temperatures ranging from 100~mK (bottom curve)
to 5.7~K (top curve).
\label{fig:6.10}
(b) The same data,
rescaled ac\-cor\-ding to \protect\Eq{eq3.testscaling}\protect\
with $\alpha = \half$,
and plotted vs. $v^{1/2} = (eV / \kb T)^{1/2}$.
The low-temperature, low-voltage data collapse onto a single
curve [linear for large $v^{1/2}$,
in agreement with \protect\Eq{eq3.limitV}\protect],
with deviations when the voltage exceeds 1~mV.
\label{fig:6.11} }
\end{figure}



\newpage \begin{figure}
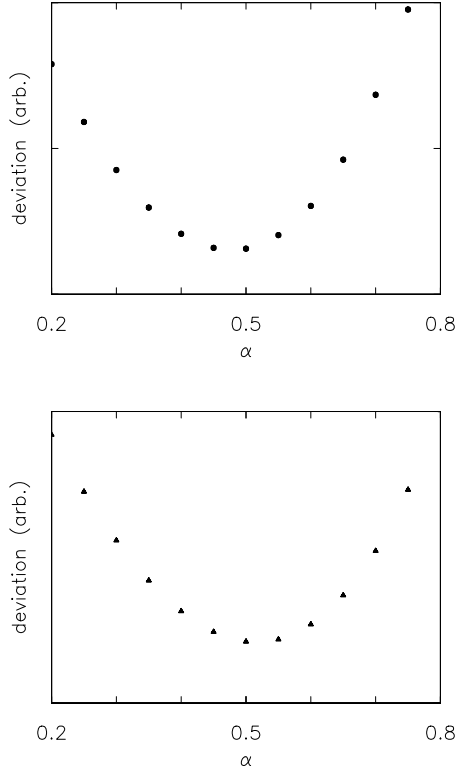

\vphantom{.}
\vspace{5cm}
\centerline{\psfig{figure=janfig9a.ps,height=6cm}}
\vphantom{.}
\vspace{-1cm}
\centerline{\psfig{figure=janfig9b.ps,height=6cm}}
\vphantom{.}
\vspace{-2cm}
\caption{
The deviation parameter $D(\alpha)$ of
\protect\Eq{eq3.dev}\protect, which quantifies the
quality of scaling, for (a) sample~\#1 and (b)
sample~\#2. The minimum of $D(\alpha)$ defines
the value of $\alpha$ that gives the best scaling,
giving $\alpha = 0.48 \pm 0.05$ for sample~\#1
and $\alpha = 0.52 \pm 0.05$ for sample~\#2.
\label{fig:deviations}}
\end{figure}

\newpage \begin{figure}
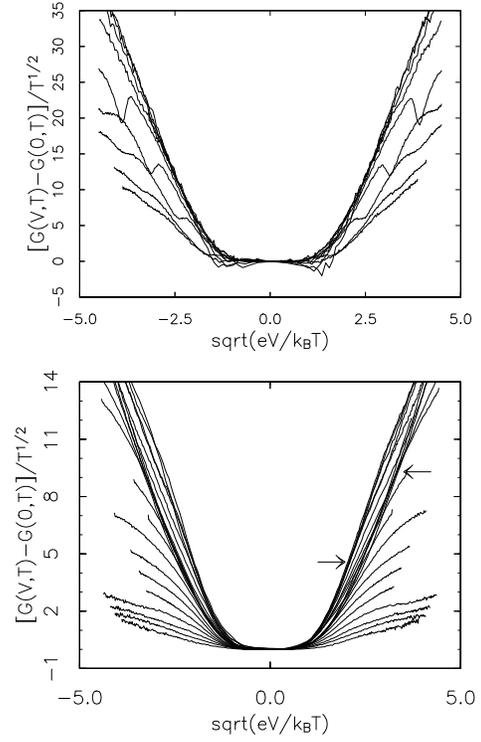

\vphantom{.}
\vspace{5cm}
\centerline{\psfig{figure=janfig10a2.ps,height=6cm}}
\vphantom{.}
\vspace{-1.5cm}
\centerline{\psfig{figure=janfig10b3.ps,height=6cm}}
\caption{%
(a) Differential conductance data for sample~\#2,
at temperatures from 200~mK to 5.7~K,
and
(b) for sample~\#3,
at temperatures from 50~mK to 7.6~K,
rescaled according to \protect\Eq{eq3.testscaling}\protect\
and plotted vs. $v^{1/2} = (eV / \kb T)^{1/2}$.
The low-voltage, low-temperature data collapse well onto one curve
for sample~\#2,
\label{fig:6.14}
but not for sample \#3, partly
due to the existence of TLSs with Kondo temperatures
within (rather than above) the  temperature range
of the measurement.
\label{fig:6.15}}
\end{figure}

\newpage \begin{figure}
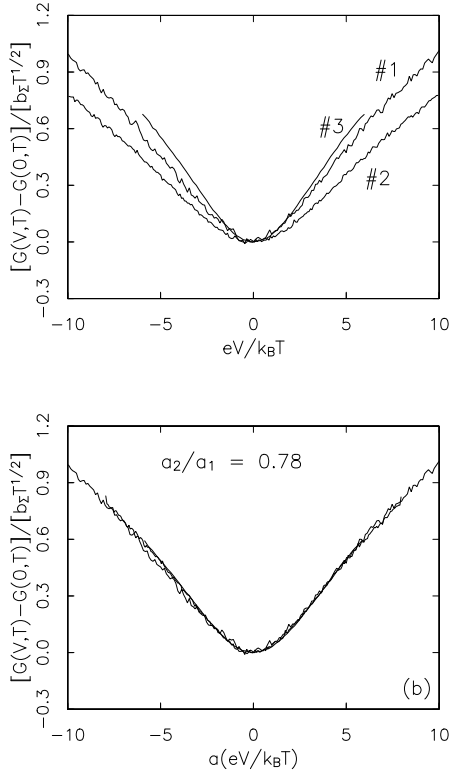

\vphantom{.}
\vspace{5cm}
\centerline{\psfig{figure=janfig11a.ps,height=6cm}}
\vphantom{.}
\vspace{-6cm}
\centerline{\psfig{figure=janfig11b.ps,height=6cm}}
\vphantom{.}
\vspace{1cm}
\caption{
Representative
conductance curves which lie along the
scaling curves for samples \#1, \#2 and \#3 of
each of the three samples in
Figs.~\protect\ref{fig:6.11}\protect(b),
\protect\fig{fig:6.14}\protect(a) and (b), respectively.
For sample~\#1, the curve
corresponds to T=1.1~K, for sample~\#2 1.4~K, and for
sample~\#3 250~mK. The reason for selecting
these particular curves was that
among all those lying along the scaling curve, they had the
best signal-to-noise ratio for each sample.
 (a) The $y$-axis is scaled by the
value of $B_{\sss \Sigma}$ determined from the temperature
dependence of the $V=0$ conductance for each sample (values
listed in Table~6.1). (b) In addition, the $x$-axis is
scaled with a number $a_i$ for each sample. The scaling curves
for all three samples seem to lie on one universal curve.
\label{fig:6.16}}
\end{figure}

 \newpage \begin{figure}
\vphantom{.}
\vspace{5cm}
\hspace{3cm}
\centerline{\psfig{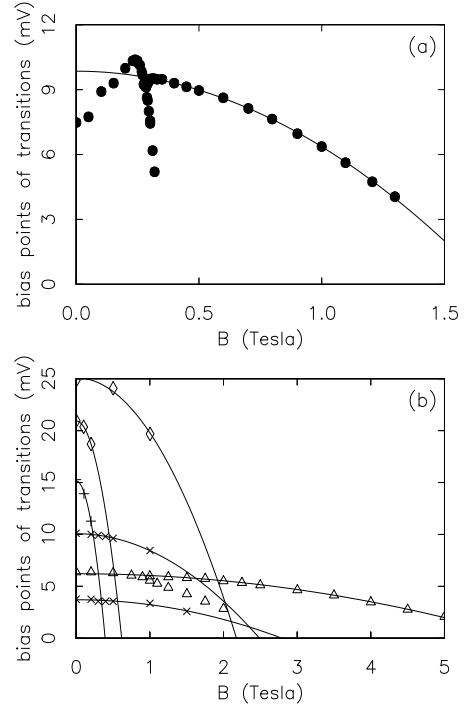}}
\vphantom{.}
\vspace{1cm}
\caption{
(a) Transition voltage $V_c (H)$ of the conductance transition 
 \protect\cite{RB95}\protect as a
function of magnetic field for a quenched Cu constriction at
4.2~K, showing bifurcation (Cu.9b,iv). At high fields $V_c (H) \to 0$
(Cu.9b,v), the dependence on $H$ being quadratic.
(b) $V_c (H)$ for five other samples at 4.2~K, with associated
decreases to 0, quadratically in $H$.
\label{fig:Vcmotion}
}
\end{figure}

\newpage \begin{figure}
\vphantom{.}
\vspace{5cm}
\centerline{\psfig{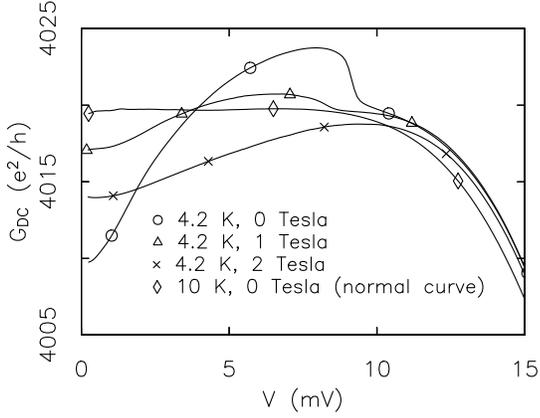}}
\vphantom{.}
\vspace{-2cm}
\caption{
The $DC$ conductance $G_{\sss DC}$ (as opposed
to differential conductance $G$ used in all other plots)
at several magnetic fields for a quenched Cu sample
 \protect\cite{RB95}\protect. 
An increasing magnetic field broadens the conductance transition
at $V_c$ and  moves $V_c$ toward zero voltage, destroying the enhancement
at $V=V_c$ of $G_{\sss DC}$ above the normal conductance.
Note that although a large applied magnetic field eliminates
the conductance transition, a $V=0$ minimum in the conductance
remains.
\label{fig:condtrans}
}
\end{figure}

\newpage \begin{figure}
\vphantom{.}
\vspace{7cm}
\centerline{\psfig{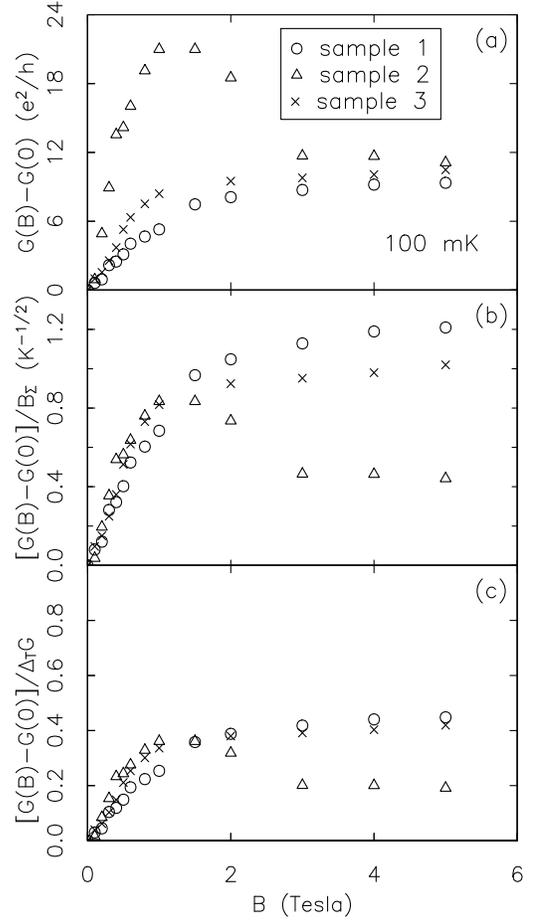}}
\vphantom{.}
\vspace{1cm}
\caption[Magnetic field dependence of the $V$=0 conductance
for the 3 unannealed Cu samples at 100~m.]{
Magnetic
field dependence of the $V$=0 conductance
for the 3 unannealed Cu samples at 100~mK. (a) Absolute
magnetoconductance. (b) Magnetoconductance scaled by
the value of $B_{\sss \Sigma}$ for each sample.
(c) Magnetoconductance relative to the change in conductance
between 100~mK and 6~K. An applied magnetic field alters, but does
not eliminate, the zero-bias conductance signal due to TLSs.
\label{fig:6.17}}
\end{figure}

\newpage \begin{figure}
\vphantom{.}
\vspace{7cm}
\centerline{\psfig{figure=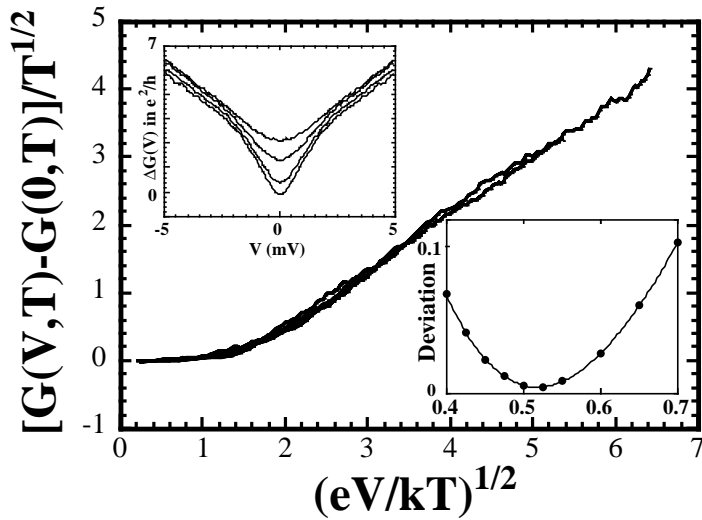,height=7cm}}
\vphantom{.}
\vspace{1cm}
\caption{The differential conductance
for a 20~$\Omega$ Ti constriction (sample~4),
for temperatures of 6.0, 4.0, 2.0 and 1.4K,  scales well 
 when plotted in the scaling
form $[G(V,T) - G(0,T)]/T^{1/2} $ vs. $(eV / \kb T)^{1/2}$
of  \protect\Eq{eq3.testscaling}\protect\,
with $\alpha = \half$ [see (Ti.2)]. Top inset: the unscaled
conductance $\Delta G(V) = G(V,T) - G(0,1.4 \mbox{K})$. Bottom inset:
The  deviation parameter $D(\alpha)$ of
\protect\Eq{eq3.dev}\protect, which quantifies the
quality of scaling, according to which $\alpha = 0.52 \pm 0.05$.}
\label{shashi1}
\end{figure}

\newpage \begin{figure}
\vphantom{.}
\vspace{2cm}
\centerline{\psfig{figure=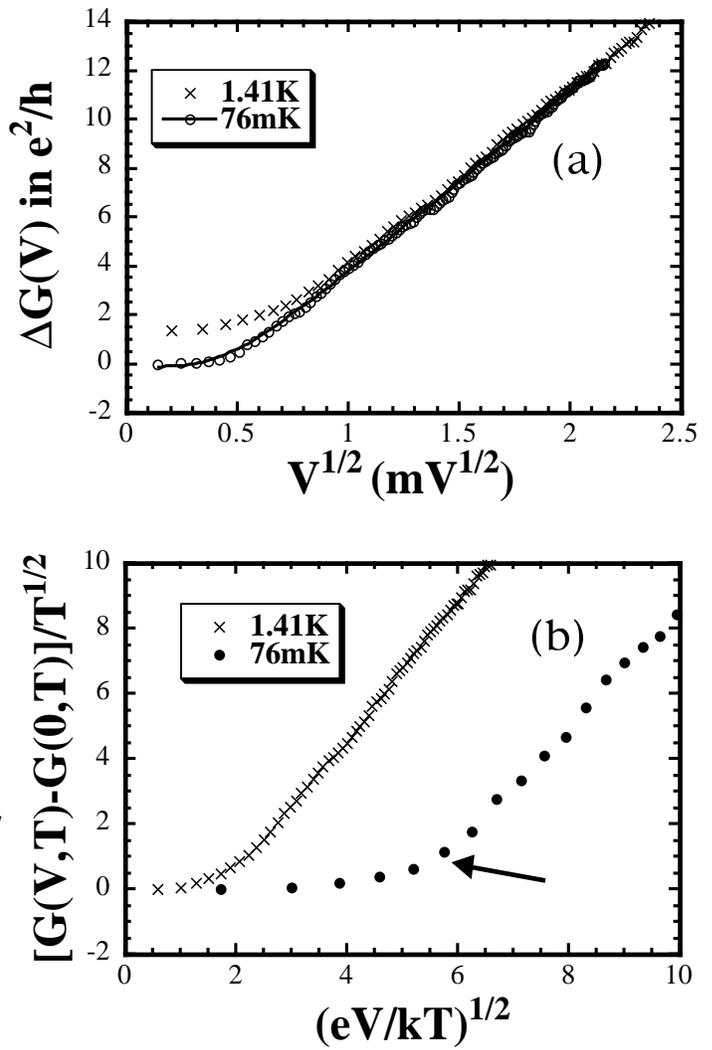,height=14cm}}
\vphantom{.}
\vspace{1cm}
\caption{(a) $\Delta G(V) = G(V,T) - G(0,0.76 \mbox{mK})$ vs. $V^{1/2}$ 
for a 19~$\Omega$ Ti constriction (sample~5) at 1.41K and 76mK.
(b) The same data do {\em not}\/ scale when plotted as 
 $[G(V,T) - G(0,T)]/T^{1/2} $ vs. $(eV / \kb T)^{1/2}$, due to
``premature saturation'' (see arrow) of the 76mK curve, as discussed
in (Ti.3). }
\label{shashi2}
\end{figure}

\newpage \begin{figure}
\vphantom{.}
\vspace{2cm}
\centerline{\psfig{figure=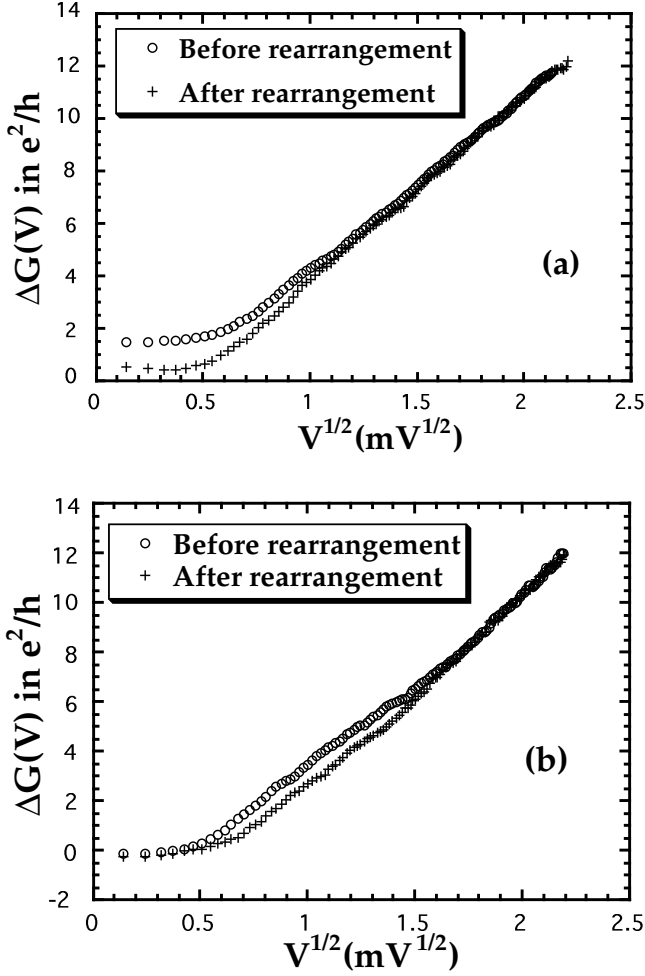,height=14cm}}
\vphantom{.}
\vspace{1cm}
\caption{(a) The ZBA of the Ti sample~5 before and after 
electromigration [see (Ti.4)], which  changed the
 saturation energy $T_x$ of Eq.~(\protect\ref{Gstressed}) from  2.3 to 1.4K.
(b) A 22~$\Omega$ Ti constriction (sample~6) shows 
two distinct saturation energies $T_{x1}$ and $T_{x2}$,
which change upon electromigration from 0.9 to 1.5K and 
9.7 to 6.8K, respectively. }
\label{shashi3}
\end{figure}

\newpage \begin{figure}
\vphantom{.}
\vspace{7cm}
\centerline{\psfig{figure=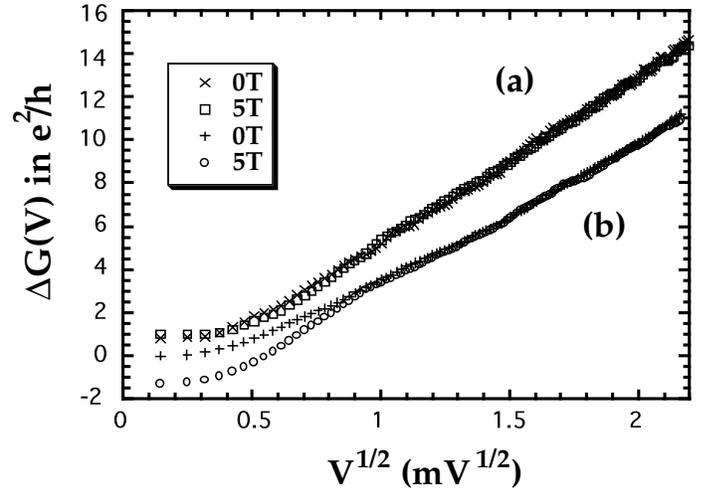,height=7cm}}
\vphantom{.}
\vspace{1cm}
\caption{The magnetic field dependence (at 0 and 5T) at 100mK for
two Ti constrictions [see (Ti.5)]: (a) sample~5; (b) sample~6. Data for (a) are
offset by $e^2/h$ at $V=0$ for clarity.}
\label{shashi4}
\end{figure}

\newpage \begin{figure}
\vphantom{.}
\vspace{7cm}
\centerline{\psfig{figure=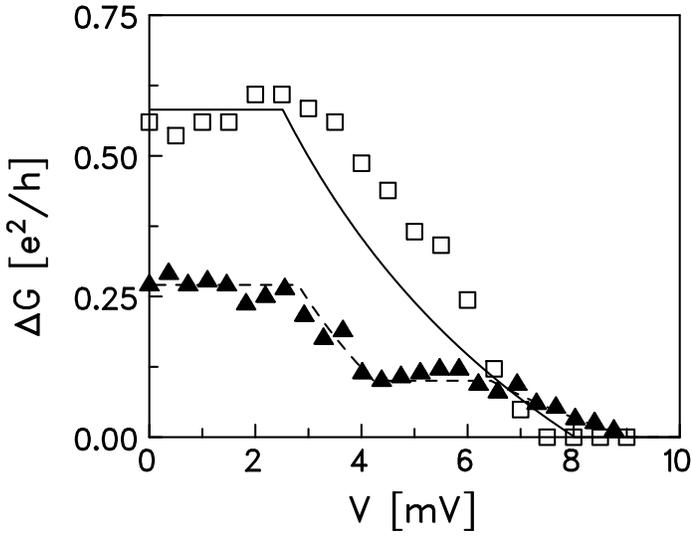,height=7cm}}
\vphantom{.}
\vspace{1cm}
\caption{ZBA fluctuations $\Delta G(V)$
[see (MG.2)] due to the presence of a slow fluctuator
in two metallic glass constrictions studied in
Ref.~\protect\cite{KSvK95}\protect.
The squares give  $\Delta G (V)$ for Fig.~2, curve 3 
of \protect\cite{KSvK95}\protect;
the triangles give the noise amplitude multiplied by 2 (for visibility) 
of  Fig.~4, curve 1 of \protect\cite{KSvK95}\protect (uncertainties 
 $\sim 0.1 e^2/h$). The fits  were obtained in
 Ref.~\protect\cite{vDZZ97}\protect\
using VZ's theory to calculate $\Delta G(V)$. For the solid curve
a single fast TLS was assumed, 
experiencing  modulations in asymmetry energy between $\Delta_{z} =
8$~meV and $\Delta'_{z} = 3$~meV, with $\Tk = 17$~K. For
the dashed curve two fast TLSs were assumed, 
with  $\Delta_{z1} = 9, \Delta_{z1}' = 6.2$~meV,  $T_{K1} = 8.9$~K, and  
 $\Delta_{z2} = 4.2, \Delta_{z2}'= 2.8$~meV,  $T_{K2} = 6.2$~K. }
\label{fig:zarand}
\end{figure}

\newpage
\begin{table}
\caption[Measured parameters of the scaling functions for the Kondo signals
in~3 Cu samples.]{
\label{tab:constants}
 Measured parameters of the scaling functions for the ZBAs
in~3 Cu constrictions (samples 1 to 3) and one Ti constriction
(sample~4). $B_{\sss \Sigma}$,
$F_o$ and $F_1$
have units $K^{-1} e^2 / h$, and $\Gamma_1$
is dimensionless.}
\begin{tabular}{|cr@{$\,\pm\,$}l
r@{$\,\pm\,$}l
r@{$\,\pm\,$}lr@{$\,\pm\,$}l|}
\hline
\# &
\multicolumn{2}{c}{$B_{\Sigma}$}        &
\multicolumn{2}{c}{$F_0$}    &
\multicolumn{2}{c}{$F_1$}    &
\multicolumn{2}{c|}{$\Gamma_1= {F_1 \over B_{\Sigma}}$}
\\ \hline
1 & 7.8 & 0.2
& 4.2 & 0.3 &--5.7 & 0.9   &--0.73 & 0.11
\\
2 & 25.2 & 0.7
& 12.8 & 0.8 &--19.7 & 1.5 &--0.78 & 0.06
\\
3 & 10.3 & 0.4  
& 6.0 & 0.6 &--7.7 & 1.6   &--0.75 & 0.16
\\
4 & 1.69 & 0.2  
& 0.89 & 0.05 &--1.37 & 0.1   &--0.81 & 0.10
\\ \hline
\end{tabular}
\end{table}

\end{document}